\documentclass[12pt]{article}
\usepackage{amsmath,amsfonts,amssymb}
\usepackage{graphicx}
\usepackage{authblk}
\usepackage[super]{natbib}

\usepackage{hyperref}
\hypersetup{
    bookmarks=true,         
    unicode=false,          
    pdftoolbar=true,        
    pdfmenubar=true,        
    pdffitwindow=false,     
    pdfstartview={FitH},    
    pdftitle={My title},    
    pdfauthor={Author},     
    pdfsubject={Subject},   
    pdfcreator={Creator},   
    pdfproducer={Producer}, 
    pdfkeywords={keyword1, key2, key3}, 
    pdfnewwindow=true,      
    colorlinks=true,       
    linkcolor=red,          
    citecolor=blue,        
    filecolor=cyan,         
    urlcolor=magenta        
}

\hypersetup{colorlinks=true,linkcolor=blue,urlcolor=blue,citecolor=blue}

\usepackage{graphicx}
\usepackage{amsmath}
\usepackage{amssymb}
\usepackage{amsbsy}
\usepackage{amsfonts}

\newcommand*\physrep{Phys. Rep.}
\newcommand*\physscr{Phys. Scr.}

\begin{document}

\title{Melnikov--Arnold integrals and \\ optimal normal forms}
\author{Ivan~I.~Shevchenko$^{1,2}$}

\maketitle

\noindent $^1$Saint Petersburg State University, 7/9
Universitetskaya nab., \\ 199034 Saint Petersburg, Russia \\
email: {\tt i.shevchenko@spbu.ru}

\noindent $^2$Institute of Applied Astronomy, Russian Academy of
Sciences, \\ 191187 Saint~Petersburg, Russia


\begin{abstract}
The Melnikov--Arnold integrals (MA-integrals) is a well-known
instrument used to measure the splitting of separatrices in
Hamiltonian systems. In this article, we explore how calculation
of MA-integrals can be used as well to estimate sizes of secondary
resonances. Within the standard map model, we show how the newly
developed MA-based procedure allows one to estimate the sizes of
secondary resonances of any order (up to the order of the optimal
normal form), without relying on the cumbersome traditional
normalization procedure.
\end{abstract}


\maketitle

\bigskip

{\bf
\noindent The Melnikov--Arnold integrals (also called
Poincar\'e--Melnikov integrals) is a well-known instrument used to
measure the splitting of separatrices in Hamiltonian systems. On
the other hand, algorithms of normalization in Hamiltonian
dynamics allow one to remove non-resonant harmonics until the
harmonics corresponding to the optimal normal form are generated.
We explore how calculations of the Melnikov--Arnold integrals can
be used to obtain resonant normal forms for a given Hamiltonian
system. The study is accomplished in the framework of the
perturbed first fundamental (pendulum) model of nonlinear
resonance. Within the standard map model, as a paradigmatic
example, we show how the proposed procedure allows one to
straightforwardly estimate the sizes of secondary resonances of
any order, up to the order of the optimal normal form. The data on
sizes of the resonances, as obtained by our MA-based method, are
all consistent with the data earlier obtained by direct
normalization. The convenience of our approach is in contrast with
traditional normalization procedures, which are usually too
cumbersome and complicated to allow one to obtain normal forms
even of low orders.
}

\section{Introduction}
\label{intro}

The Melnikov--Arnold integrals (MA-integrals) is a well-known
instrument used to explore chaotic dynamics in
Hamiltonian systems: it is used to analytically calculate the
amplitudes of separatrix splitting \cite{Morbi02}; it emerges in
derivation of separatrix maps \cite{C79PhR,S20ASSL}. In this
article, we explore how calculation of MA-integrals can be
used as well to estimate sizes of secondary resonances.

Following Ref.~\cite{MG97PhyD}, let us consider the Hamiltonian

\begin{equation}
{\cal H}(p,\varphi,t)={\cal H}_{\rm res}(p,\varphi) + {\cal R}(p,\varphi,t) ,
\label{ex-split}
\end{equation}

\noindent where ${\cal H}_{\rm res}$ is the non-perturbed pendulum
Hamiltonian:

\begin{equation}
{\cal H}_{\rm res}(p,\varphi)=\frac{p^2}{2} - \epsilon \cos \varphi ;
\label{ex-split_1}
\end{equation}

\noindent and ${\cal R}$ is the perturbation:

\begin{equation}
{\cal R}(p,\varphi,t)=-\frac{\epsilon\mu}{2}
\{\cos[(k+1)\varphi-t]+\cos[(k-1)\varphi-t]\} ;
\label{ex-split_2}
\end{equation}

\noindent the variables $p$ and $\varphi$ are conjugate canonical
momenta and coordinates, $t$ is time; $\epsilon$ and $\mu$ are
constant parameters; integer $k>0$.

It is convenient to introduce designations

\begin{equation}
\varepsilon = \frac{\mu}{2}, \quad
\lambda = \frac{1}{\sqrt{\epsilon}},
\label{vareps}
\end{equation}

\noindent for, respectively, the relative strength of perturbation
and the ``adiabaticity parameter,'' which is discussed further on.

According to Ref.~\cite{Morbi02} (p.~87), for the perturbation
harmonics that are resonant at the border of the domain spanned by
the pendulum separatrices, the given Hamiltonian
(\ref{ex-split})--(\ref{ex-split_2}) can be considered a
paradigm for the optimal normal form in the perturbed pendulum
problem.\footnote{The perturbed nonlinear pendulum serves as the
first fundamental model of perturbed nonlinear resonance; see
Refs.~\cite{S20ASSL,S24JETPL}.}

The harmonics of the perturbation ${\cal R}$ are resonant at the
borders of the zone

\begin{equation}
|p| < 2\sqrt{\epsilon}
\label{p_crit}
\end{equation}

\noindent if

\begin{equation}
k \approx 1/(2\sqrt{\epsilon}) ,
\label{k_MG}
\end{equation}

\noindent see Ref.~\cite{Morbi02}; i.~e., at the borders of the
zone $ |p| < 2/\lambda $ if $k \approx \lambda / 2$, in our
notations. Zone~(\ref{p_crit}) is circumscribed by the
separatrices of the ${\cal H}_{\rm res}$ pendulum; therefore, the
given harmonics correspond to secondary resonances, as defined in
Refs.~\cite{MG97PhyD,Morbi02}.

Normalization algorithms allow one to remove non-resonant
harmonics until the harmonics corresponding to the optimal normal
form are generated \cite{MG97PhyD,Morbi02}. The sizes of such
harmonics are $O \left( (\epsilon\mu)^{ \left| k -
\frac{\lambda}{2} \right|} \right)$. In Ref.~\cite{MG97PhyD}, it
is argued that the separatrix splitting amplitude is generically
determined by the value of the coefficient of the ``leading
secondary resonance,'' i.~e., by the remainder ${\cal R}$ of the
optimal normal form.

The normalization order given, in our terms, by Eq.~(\ref{k_MG}),
corresponds to the optimal normal form, as determined in
Ref.~\cite{MG97PhyD}. In what follows, this assertion
($k_\mathrm{optimal} \approx \lambda / 2$) is called the
Morbidelli--Giorgilli prescription (MG-prescription).

Note that, further on, we adopt a more general definition for the
secondary resonances; namely, in the considered
model~(\ref{ex-split}), by the secondary resonances we imply any
resonances that arise due to interaction of the main resonances
explicitly present in the initial (yet non-normalized)
Hamiltonian.

In this article, we aim to explore how calculation of the
MA-integrals can be used to estimate sizes of the secondary
resonances. To this end, we analytically calculate the
MA-integrals in dependence on the order $k$ and the adiabaticity
parameter $\lambda$.

As a paradigmatic example, within the standard map model, we show
how the newly developed MA-based procedure allows one to
straightforwardly estimate the sizes of secondary resonances of
any order (up to the order of the optimal normal form), in
contrast with traditional normalization procedures, which are
usually too cumbersome and complicated to allow one to obtain
normal forms even of low orders.

First, let us consider main analytical instruments that we use in
what follows.

\section{The separatrix map}
\label{sec_sm}

Let us take, for the model of perturbed nonlinear resonance, a
perturbed pendulum Hamiltonian, which is rather similar, although
not identical, to Eqs.~(\ref{ex-split})--(\ref{ex-split_2}):

\begin{equation}
H = \frac{{\cal G} p^2}{2} - {\cal F} \cos \varphi +
    a \cos(k \varphi - \tau) + b \cos(k \varphi + \tau) .
\label{h}
\end{equation}

\noindent Its first two terms represent the unperturbed part:

\begin{equation}
H_0 = \frac{{\cal G} p^2}{2} - {\cal F} \cos \varphi , \label{h0}
\end{equation}

\noindent which is the same as ${\cal H}_{\rm res}$ in
Eq.~(\ref{ex-split_1}), i.~e., it is the unperturbed pendulum with
angle $\varphi$ (in the pendulum model of resonance, it is the
resonant phase angle); $p$ is the conjugated canonical momentum.
The last two terms in Eq.~(\ref{h}) represent a periodic
perturbation; $\tau = \Omega t + \tau_0$ is the perturbation phase
angle, $\Omega$ is the perturbation frequency, $\tau_0$ is the
initial perturbation phase. The parameters ${\cal F}$, ${\cal G}$,
$a$, $b$, and $k$ are constants; $k$ is integer. Without loss of
generality we set ${\cal G} = 1$, as adopted in
Refs.~\cite{C79PhR,MG97PhyD}.

The motion near the separatrix of Hamiltonian~(\ref{h}) with $k=1$
and $a = b$ was considered by Chirikov~\cite{C79PhR}. He showed
that it is described by a map, which he called a ``whisker map.''
Now the term ``separatrix map'' is customary for maps of this
kind. The separatrix map in Chirikov's~\cite{C79PhR} form is given
by

\begin{eqnarray}
& & w_{i+1} = w_i - W \sin \tau_i,  \nonumber \\
& & \tau_{i+1} = \tau_i +
                 \lambda \ln \frac{32}{\vert w_{i+1} \vert}
                 \ \ \ (\mbox{mod } 2 \pi),
\label{sm}
\end{eqnarray}

\noindent where $w$ denotes the relative (with respect to the
separatrix value) pendulum energy $w = \frac{H_0}{{\cal F}} - 1$;
$\tau$ is the perturbation phase; $\lambda$ and $W$ are constant
parameters; $i$ is the iteration step.

The quantity $\lambda$ is the adiabaticity parameter, already
defined above in Eq.~(\ref{vareps}). Equivalently, in the current
designations, it is

\begin{equation}
\lambda = \frac{\Omega}{\omega_0} , \label{lambda}
\end{equation}

\noindent where $\Omega$ is the perturbation frequency, and
$\omega_0 = ({\cal F G})^{1/2}$ is the frequency of
small-amplitude oscillations on resonance.

If $k=1$ and $a = b$, then

\begin{equation}
W(\varepsilon, \lambda) =
\varepsilon \lambda \cdot \left( A_2(\lambda) + A_2(-\lambda) \right) =
4 \pi \varepsilon \frac{\lambda^2}{\sinh \frac{\pi \lambda}{2}} ,
\label{W1}
\end{equation}

\noindent where the perturbation relative strength $\varepsilon =
a / {\cal F} = b / {\cal F}$, and

\begin{equation}
A_2(\lambda) = 4 \pi \lambda \frac{\exp (\pi \lambda / 2 )}{\sinh
(\pi \lambda)};
\label{A2}
\end{equation}

\noindent see Refs.~\cite{C79PhR,LL92,S00JETP,S20ASSL}.

The separatrix map theory provides a straightforward analytical
description of the structure of the near-separatrix phase space
\cite{S99CM,S20ASSL}, it allows one to locate resonances and chaos
borders in phase space, to predict emergence of marginal
resonances; see Refs.~\cite{S98PS,S20ASSL}. The separatrix map can be
linearized in $w$ to give the standard map; then the chaotic layer
borders can be located~\cite{C79PhR} by the condition of
criticality of the stochasticity parameter of the obtained
standard map.

Formula~(\ref{W1}) differs from that given in
Refs.~\cite{C79PhR,LL92} by the addend $A_2(-\lambda)$, which is
small if $\lambda \gtrsim 1$. Conversely, if $\lambda \ll 1$, its
contribution becomes significant \cite{S98PS}.

If $k \neq 1$, then $W(\varepsilon, \lambda)$ is generalized to

\begin{equation}
W(\varepsilon, \lambda) = \varepsilon \lambda \cdot \left(
A_{2k}(\lambda) + A_{2k}(-\lambda) \right) ,
\label{W_k}
\end{equation}

\noindent where $\varepsilon = a / {\cal F} = b / {\cal F}$; see
Refs.~\cite{S00JETP,S20ASSL}.

If the perturbation is asymmetric, i.~e., $a \neq b$, then the
separatrix map differs from representation~(\ref{sm}), because
energy increments are different for the prograde and retrograde
motions of the model pendulum.

Therefore, the separatrix map in the asymmetric case represents
the {\it separatrix algorithmic map} \cite{S99CM,S20ASSL}. The
essence of this algorithm consists of taking into account
alternations of values of the parameter $W$; they alternate when
the motion direction changes. If $k=1$, the map is given by

\begin{eqnarray}
& & \mbox{if } w_i < 0 \mbox{ and } W = W^-
              \mbox{ then } W := W^+, \nonumber \\
& & \mbox{if } w_i < 0 \mbox{ and } W = W^+
              \mbox{ then } W := W^- , \nonumber \\
& & w_{i+1} = w_i - W \sin \tau_i,   \nonumber \\
& & \tau_{i+1} = \tau_i + \lambda \ln \frac{32 }{ \vert w_{i+1} \vert }
                   \ \ \ (\mbox{mod } 2 \pi) ,
\label{sam}
\end{eqnarray}

\noindent where $\lambda$ is still given
by Eq.~(\ref{lambda}), and the parameter $W$, present in
map~(\ref{sm}), is generalized now to

\begin{eqnarray}
 & & W^+ (\varepsilon, \lambda, \eta) = \varepsilon
 \lambda \left( A_{2}(\lambda) + \eta A_{2}(-\lambda) \right), \nonumber \\
 & & W^- (\varepsilon, \lambda, \eta) = \varepsilon
 \lambda \left( \eta A_{2}(\lambda) + A_{2}(-\lambda) \right),
\label{samWpm}
\end{eqnarray}

\noindent where $\varepsilon = a / {\cal F}$, and the perturbation
asymmetry parameter $\eta = b / a$.

If $k \neq 1$, then $W^\pm (\varepsilon, \lambda, \eta)$ are generalized to

\begin{eqnarray}
 & & W_k^+ (\varepsilon, \lambda, \eta) = \varepsilon
 \lambda \left( A_{2k}(\lambda) + \eta A_{2k}(-\lambda) \right) ,
 \nonumber \\
 & & W_k^- (\varepsilon, \lambda, \eta) = \varepsilon
 \lambda \left( \eta A_{2k}(\lambda) + A_{2k}(-\lambda) \right) .
\label{samWpmk}
\end{eqnarray}

One should be aware that the expression for the increment of the
phase $\tau$ in map~(\ref{sm}) is an approximation, valid
asymptotically when the near-separatrix rotation/oscillation
period on resonance tends to infinity. Therefore, it is applicable
at low perturbation strengths (i.~e., when $W \ll 1$). Conversely,
if the perturbation is not weak, one may improve
\cite{S98PS,S20ASSL} the performance of the map by means of
replacing the logarithmic approximation of the increment by the
original exact expression through elliptic integrals:

\begin{equation}
\Delta_{i+1} \tau =
\begin{cases}
2 \lambda {\mathbf K} \left( (1+{w_{i+1} \over 2})^{1/2} \right), &
\text{if $w_{i+1} < 0$}, \\
2 \lambda (1+{w_{i+1} \over 2})^{-1/2} {\bf K} \left( (1+{w_{i+1}
\over 2})^{-1/2} \right), & \text{if $w_{i+1} > 0$},
\end{cases}
\label{dtau}
\end{equation}

\noindent where ${\mathbf K}(k)$ is the elliptic integral of the first
kind. The first line in Eq.~(\ref{dtau}) corresponds to the model
pendulum libration, and the second one to its rotation.

\section{Recurrent relations for the MA-integrals}
\label{sec_MA}

Generally, the Melnikov--Arnold integrals (also sometimes called
Poincar\'e--Melnikov integrals) provide measure of the separatrix
splitting in dynamical systems; for their general definition and
theory see Ref.~\cite{Morbi02}. In model~(\ref{h}), the
MA-integral of order $i$, as defined in Ref.~\cite{C79PhR}, is
given by the formula

\begin{equation}
 A_i(\lambda) =
 \int_{-\infty}^{\infty}\!
 \cos\left({i \over2 }\varphi(t) - \lambda t \right)\,{dt},
\label{An}
\end{equation}

\noindent where $\lambda$ is any real, and $i \geq 0$ is generally
also real (but we assume it to be integer); and

\begin{equation}
 \varphi(t) = 4 \arctan \exp (t) - \pi .
\end{equation}

\noindent Formula~(\ref{An}) can be recast as

\begin{equation}
 A_i(\lambda) =
 2 \int _{0}^{\infty}\!\cos(i \arctan \sinh (t) - \lambda t)\,{dt}.
\end{equation}

\noindent see Ref.~\cite{S00JETP}.

The MA-integrals of any order in model~(\ref{h}) can be calculated
using recurrent relations

\vspace{-4mm}

\begin{eqnarray}
 A_0(\lambda) & = & 0, \cr
 A_1(\lambda) & = & {\alpha^c_1(\lambda) + \alpha^s_1(\lambda) \over 2}, \cr
 \cdots \cr
 A_{i+1}(\lambda) & = & \frac{2 \lambda}{i} A_i(\lambda) -
                A_{i-1}(\lambda)
\label{Ana}
\end{eqnarray}

\noindent (see eqs.~(A.10) in Ref.~\cite{S00JETP}), where

\vspace{-4mm}

\begin{equation}
 \alpha^c_1(\lambda) =
 {2 \pi \over \cosh {\pi \lambda \over 2}},
\label{ac1}
\end{equation}

\begin{equation}
 \alpha^s_1(\lambda) =
 {2 \pi \over \sinh{\pi \lambda \over 2}} .
\label{as1}
\end{equation}

\noindent As soon as $A_0$ and $A_1$ are specified, all $A_i$ can
be found by formulas~(\ref{Ana}).

Let us introduce a two-dimensional map, defined by the equations

\vspace{-5mm}
\begin{eqnarray}
u_{i+1} &=& A_{i+1} - A_i , \nonumber \\
v_{i+1} &=& A_{i+1} ,
\label{MA_map1}
\end{eqnarray}

\noindent or, equivalently,

\vspace{-5mm}
\begin{eqnarray}
u_{i+1} &=& 2 \left( \frac{\lambda}{i} - 1 \right) v_i + u_i , \nonumber \\
v_{i+1} &=& v_i + u_{i+1} ,
\label{MA_map}
\end{eqnarray}

\noindent with the initial data $v_1 = A_1$, $u_1 = A_1$.
Note that the explicit dependence of the right-hand side
of map~(\ref{MA_map}) on the iteration index
(abstract ``time'') $i$ can be removed by
introducing a third variable and making the map three-dimensional.

\begin{figure}[th!]
\centering
\includegraphics[width=0.5\textwidth]{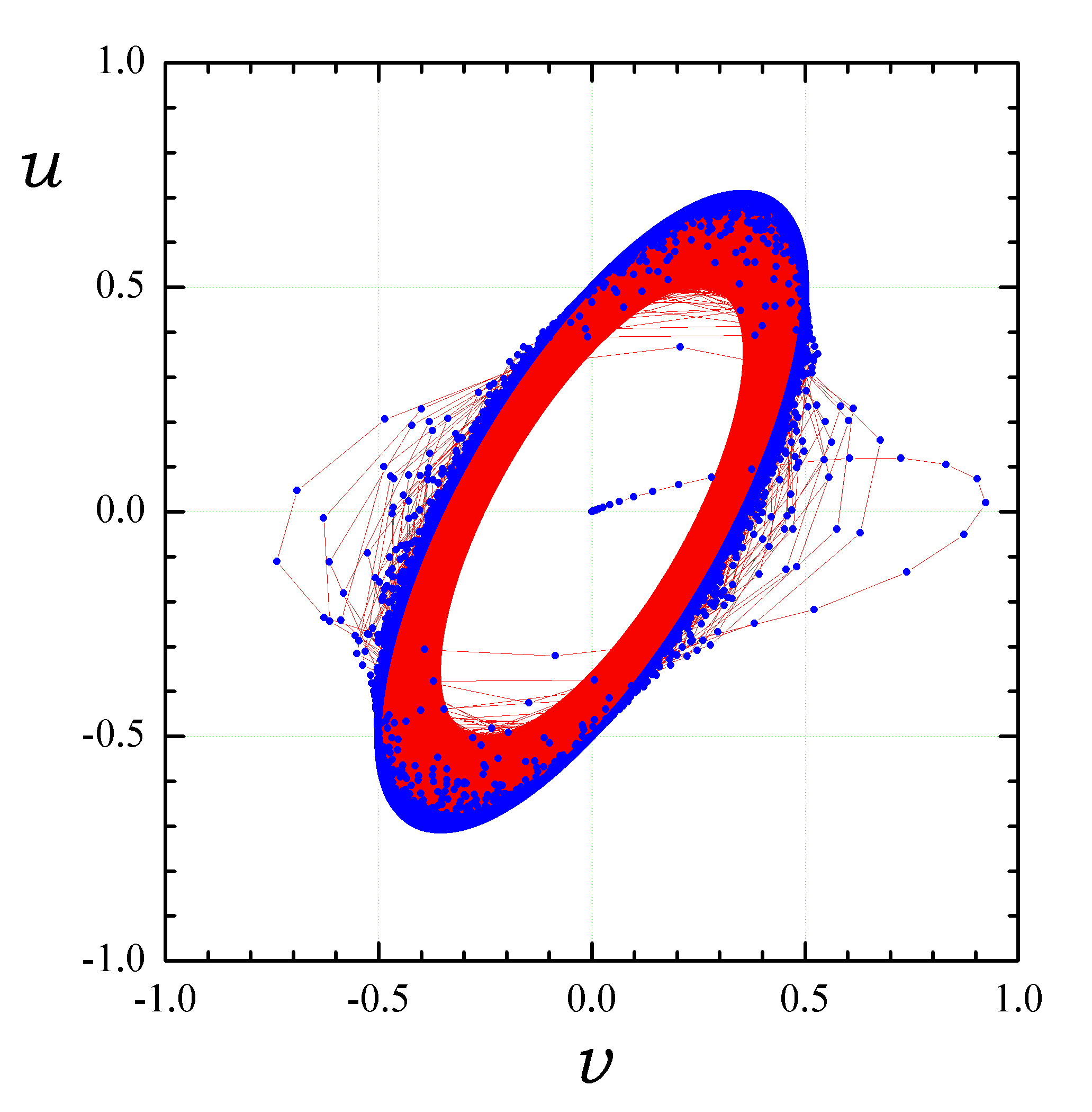} \\
\caption{The ${\cal O}$-attractor of map~(\ref{MA_map}), at
$\lambda = 100$. Each iteration is shown by a blue dot; the dots
are connected by thin red lines, to trace the evolution with
increasing $i$. The iterations start at the graph centre.}
\label{la_100}
\end{figure}

In Fig.~\ref{la_100}, we present an example of the map
behaviour, at $\lambda = 100$; the number of iterations is 50000.
Each iteration is shown by a blue dot; the dots are connected by
thin red lines, to trace the evolution with increasing $i$. The
evolution starts close to the centre, and then, at large $i$,
goes around an oval.
Obviously, an $O$-like attractor exists in the phase plane.

This behaviour is typical if $\lambda \gtrsim 1$. In this case,
the second addend in expression~(\ref{W1}) for the parameter $W$
can be ignored. Conversely, if $\lambda \ll 1$, this addend
becomes significant, and one should take it into account. Then, in
the corresponding phase portrait (not shown here) an $X$-like
(cross-like) attractor emerges, instead of the $O$-like one. We do
not provide details here, since explorations of the properties of
such maps are beyond the scope of our present study.

\section{Amplitudes of the separatrix splitting}

First of all, let us reconstruct the graphs in figures~4--5 in
Ref.~\cite{MG97PhyD} and in figure~4.7 in Ref.~\cite{Morbi02};
these graphs represent the amplitudes of the separatrix splitting
as functions of $k$ and $\mu$. In Ref.~\cite{MG97PhyD}, they were
built by measuring the splitting in direct numerical computations.
In the ``Conclusions and discussion'' section of
Ref.~\cite{MG97PhyD}, the authors write ``... it could be
interesting to use the method of Melnikov's integral in order to
compute the splitting of separatrices for the
Hamiltonian~(12)\footnote{In our article, it is
Hamiltonian~(\ref{ex-split})--(\ref{ex-split_2}).}.'' This is just
what we perform in what follows, calculating the MA-integrals in
the given problem analytically.

Note that, in Eq.~(\ref{ex-split_1}), the sign at the second term
of $H_0$ is chosen to be minus, as adopted in
Ref.~\cite{MG97PhyD}; therefore, the stable position of the
pendulum is at $\varphi = 0$, and the unstable one is at $\varphi
= \pi$. In Ref.~\cite{C79PhR}, the sign is different (plus);
therefore, the stable position is at $\varphi = \pi$, and the
unstable one is at $\varphi = 0$. This is important to take into
account in what follows, because the splitting size depends on
the value of $\varphi$ where it is measured.

It is necessary to be able to convert the splitting size, given in
the relative energy $w = \frac{H_0}{{\cal F}} - 1$ (in which the
separatrix map~(\ref{sm}) operates), into the splitting size in
the canonical momentum $p$ in Hamiltonian~(\ref{h}). We use the
relations:

\begin{equation}
\Delta p_{\varphi = 0} \approx \frac{1}{2} \omega_0 \Delta w =
\frac{\Delta w}{2 \lambda} , \quad
\Delta p_{\varphi = \pi}\approx \omega_0 \sqrt{\Delta w} =
\frac{\sqrt{\Delta w}}{\lambda} ,
\label{Delta_p_w}
\end{equation}

\noindent valid at $\varphi = 0$ and $\varphi = \pi$,
respectively. Here, in accordance with the used expression for the
Hamiltonian, we set the perturbation frequency $\Omega = 1$.

Relations~(\ref{Delta_p_w}) are derivable from the Hamiltonian equation

\begin{equation}
\frac{\partial H_0}{\partial p} ( = p) = \dot \varphi ,
\end{equation}

\noindent where $H_0$ is given by Eq.~(\ref{h0}), and ${\cal F} =
\omega_0^2$. As soon as $w = \frac{H_0}{\omega_0^2} - 1$, one has

\begin{equation}
\frac{\partial w}{\partial p} = \frac{1}{\omega_0^2}
\frac{\partial H_0}{\partial p} .
\end{equation}

\noindent Substantiating the differentials, one arrives at

\begin{equation}
\frac{\Delta w}{\Delta p} \approx \frac{p}{\omega_0^2} .
\end{equation}

\noindent At $\varphi = 0$, the splitting is calculated at the
separatrix extrema in $p$, where $p = \pm 2 \omega_0$. Therefore,

\begin{equation}
\Delta p \approx \frac{\omega_0}{2} \Delta w = \frac{\Delta w}{2 \lambda} ,
\label{dp_0}
\end{equation}

\noindent giving the first one of relations~(\ref{Delta_p_w}).

At $\varphi = \pi$, the splitting is calculated at the unstable
point; here $p \approx \Delta p$. Hence

\begin{equation}
\Delta p \approx \omega_0 \sqrt{\Delta w} = \frac{\sqrt{\Delta w}}{\lambda} ,
\label{dp_pi}
\end{equation}

\noindent giving the second one of relations~(\ref{Delta_p_w}).

In paradigm~(\ref{ex-split})--(\ref{ex-split_2}), in the case of
the model pendulum prograde motion, one finds

\begin{equation}
\Delta w = W_{k+1}^+(\varepsilon,\lambda,\eta) +
W_{k-1}^+(\varepsilon,\lambda,\eta),
\label{Delta_w_W}
\end{equation}

\noindent where $W_m^+$ at $m=k+1$ and $k-1$ are given by
formula~(\ref{samWpmk}), and $\eta = 0$.

\begin{figure}[th!]
\centering
\includegraphics[width=0.5\textwidth]{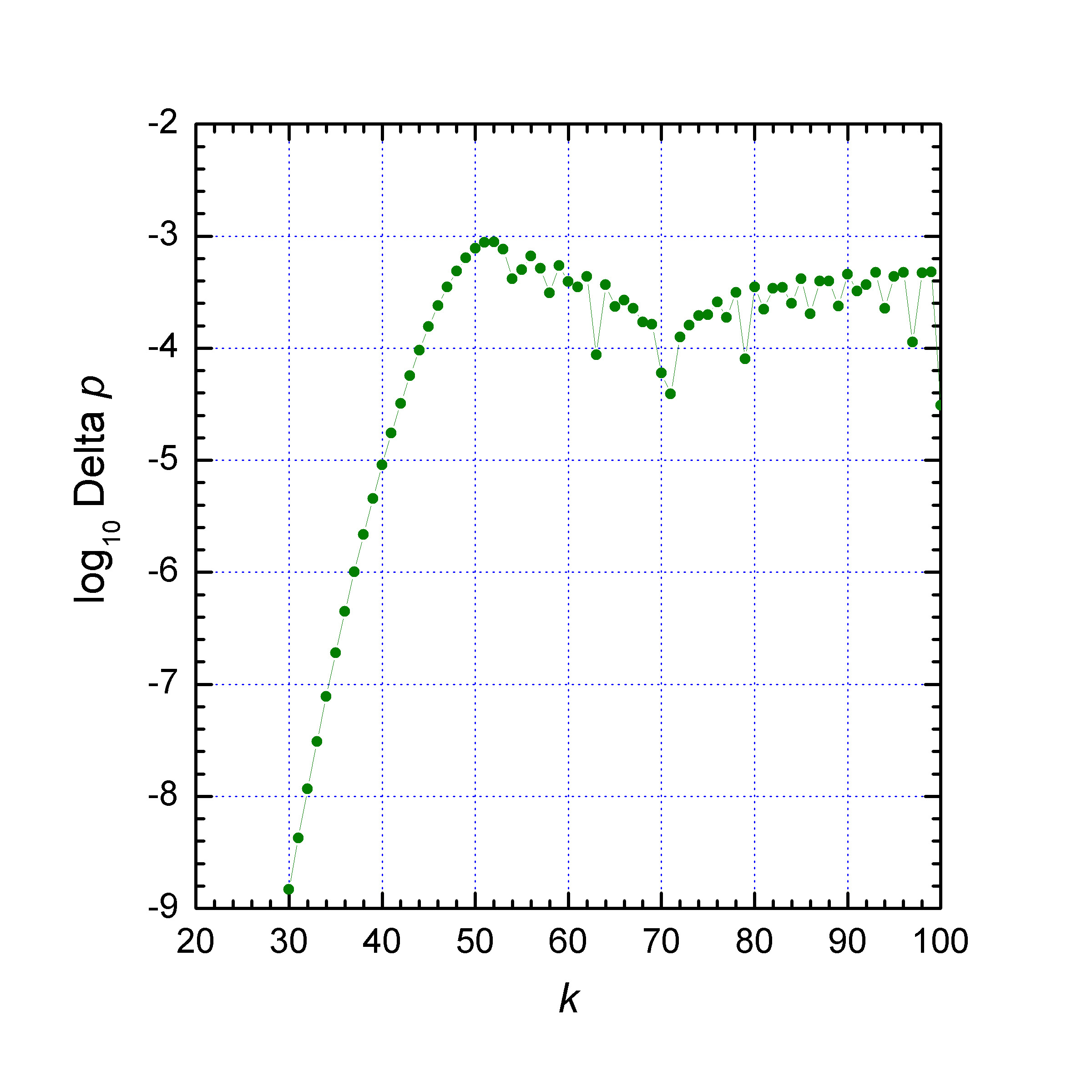} \\
\caption{The separatrix splitting amplitude in
model~(\ref{ex-split})--(\ref{ex-split_2}), as a function of $k$.
The splitting is calculated at $\varphi = \pi$.}
\label{MSPX_qpi_100}
\end{figure}

In Ref.~\cite{MG97PhyD}, the splitting amplitude is numerically
measured as the distance $\Delta p$ between the stable and
unstable manifolds at $\varphi=\pi$. Here we obtain the splitting
size analytically.
Specifically, at $\varphi = \pi$ in
model~(\ref{ex-split})--(\ref{ex-split_2}), using
Eq.~(\ref{dp_pi}), we calculate

\begin{equation}
\Delta p = \frac{1}{\lambda} \cdot \left(
W_{k+1}^+(\varepsilon,\lambda,0) +
W_{k-1}^+(\varepsilon,\lambda,0) \right)^{1/2} ,
\label{Delta_p_Wk}
\end{equation}

\noindent and, in Fig.~\ref{MSPX_qpi_100}, we present the
calculated splitting size as a function of $k$, for the same
values of parameters as used in figures~4--5 in
Ref.~\cite{MG97PhyD}; i.~e., we set $\lambda=100$ and $\varepsilon
= 0.00005$ (note that $\varepsilon = \mu/2$). Thus,
Fig.~\ref{MSPX_qpi_100} is an analytically constructed analogue of
the numerically constructed figure~5 in Ref.~\cite{MG97PhyD} (also
reproduced as figure~4.7b in Ref.~\cite{Morbi02}). The graph
obtained here turns out to ideally resemble figure~5 in
Ref.~\cite{MG97PhyD}, although the methods to build them are
totally different: analytical one (here) and numerical one (in
Ref.~\cite{MG97PhyD}), respectively.

A notable pattern in Fig.~\ref{MSPX_qpi_100} is a prominent dip at
$k \sim 70$. As soon as we calculate the splitting analytically,
the reason for its appearance can be straightforwardly identified.
It turns out that it is caused by a ``beating'' of two addends,
comprising the MA-integral here. Indeed, the
Hamiltonian~(\ref{ex-split})--(\ref{ex-split_2}) contains two
perturbation terms, the ``$k-1$'' and ``$k+1$'' ones; they produce
these two addends. Their beating leads to $\Delta w$ (given by
Eq.~(\ref{Delta_w_W})) sharply decreasing in a neighbourhood of
$\lambda=100$, when $k \sim 70$.

\begin{figure}[th!]
\centering
\includegraphics[width=0.5\textwidth]{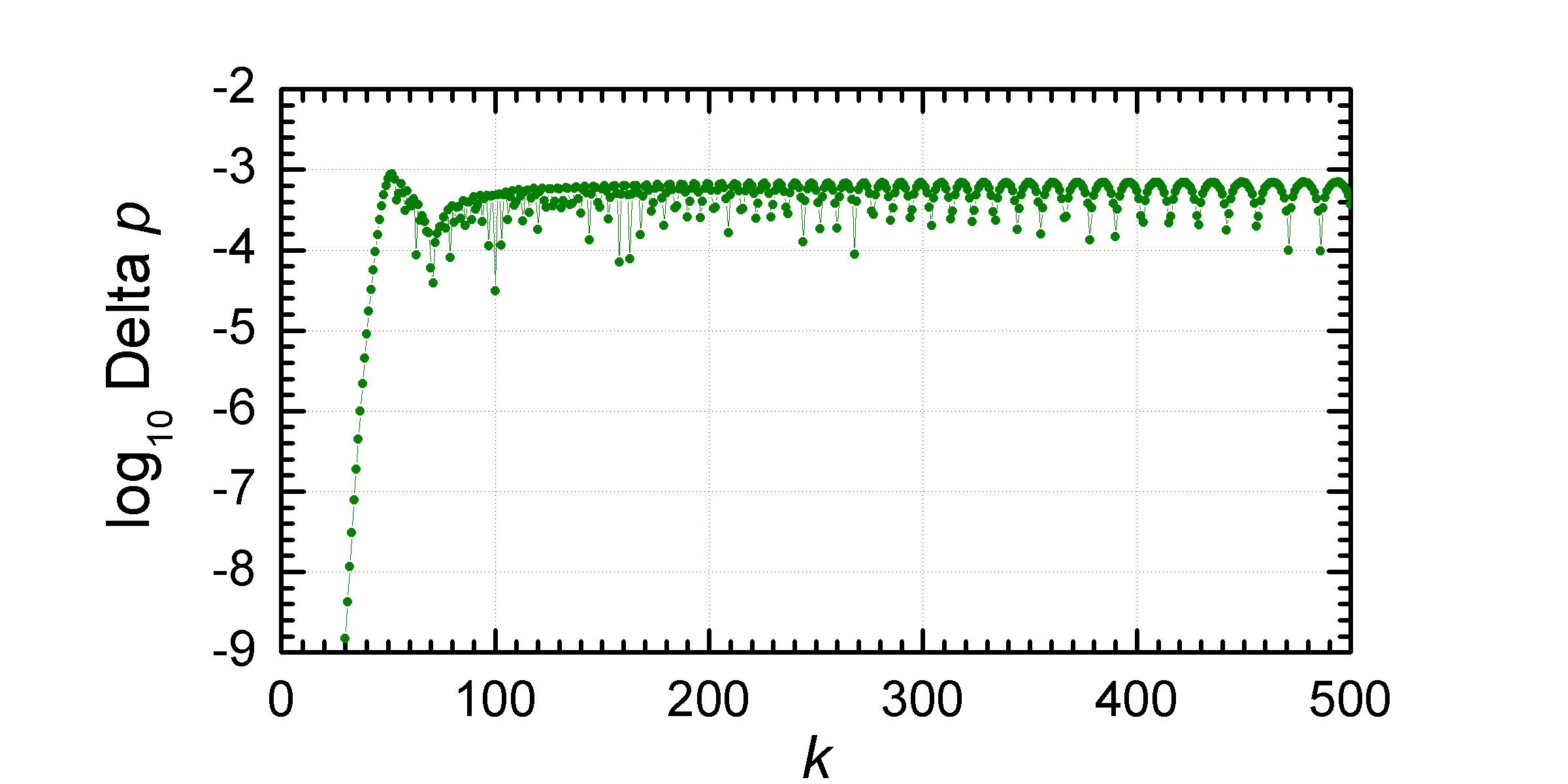} \\
\caption{The same as Fig.~\ref{MSPX_qpi_100}, but in a broader range of $k$.}
\label{MSPX_qpi_500}
\end{figure}

Our Fig.~\ref{MSPX_qpi_500} is the same as
Fig.~\ref{MSPX_qpi_100}, but the calculated data are presented in
a broader range of $k$. Here we find another notable pattern,
arising recurrently and prominently at $k > 100$. It resembles a
``comb'' with quasiperiodic (in $k$) vertical cogs. This comb-like
pattern arises because the MA-integral goes to zero at suitable
combinations of $k$ and $\lambda$ values. Indeed, the formula for
the MA-integral for
Hamiltonian~(\ref{ex-split})--(\ref{ex-split_2}) turns out to
contain, as a co-product, a polynomial of degree $2k$ in
$\lambda$. This polynomial has $k$ real positive roots. Thus, for
any $k>1$, there exist $k$ positive values of $\lambda$ at which
$W=0$. The recurrent pattern arises because, when $k$ is
monotonously increased, specific values of $k$ at which the
integral $W$ is close to (although not exactly) zero, are
recurrently encountered.

For example, if $k=5$, the expression for $W$ incorporates a
polynomial of degree 10 in $\lambda$:

\begin{eqnarray}
& & W(\lambda) = (4 \pi \mu/155925) \cdot (\sinh(\pi \lambda / 2)
+ \cosh(\pi \lambda / 2)) \lambda^2 \cdot
\nonumber \\
& & (2 \lambda^{10}-220 \lambda^8+7986 \lambda^6-102520 \lambda^4+ 391787 \lambda^2-277110)/ \nonumber \\
& & \sinh(\pi \lambda ) .
\label{W_example}
\end{eqnarray}

\noindent This is an exact analytical representation, as provided
by recurrent relations~(\ref{Ana}). It is clear that, if one
explores the behaviour of $\Delta p$ as a function of $\lambda$ at
a fixed $k$, then, on increasing $\lambda$ monotonously, {\it
exact} zeros for $W$ will be encountered recurrently.

\begin{figure}[th!]
\centering
\includegraphics[width=0.5\textwidth]{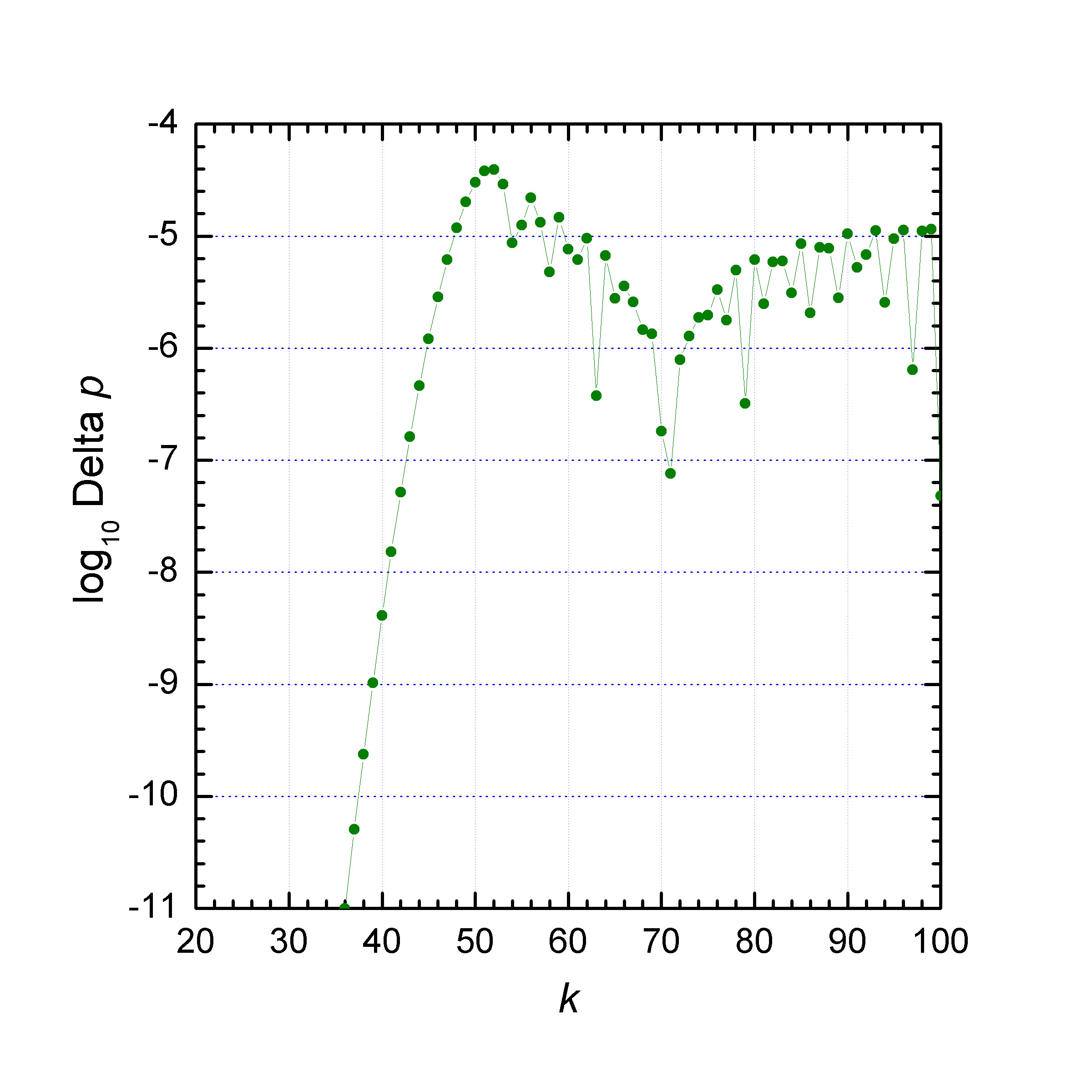} \\
\caption{The same as Fig.~\ref{MSPX_qpi_100}, but the splitting is
calculated at $\varphi = 0$.} \label{MSPX_q0_100}
\end{figure}

\begin{figure}[th!]
\centering
\includegraphics[width=0.5\textwidth]{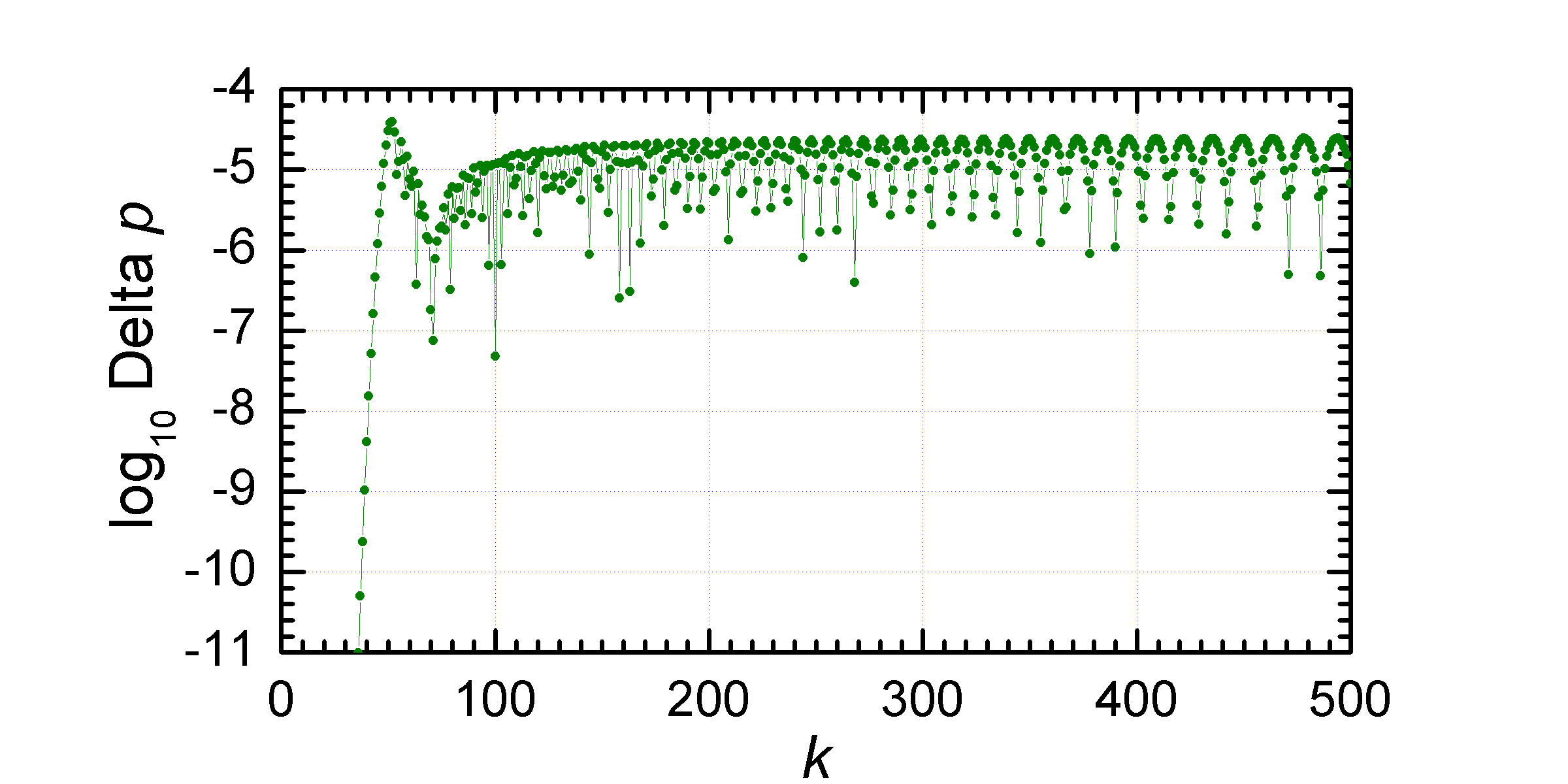} \\
\caption{The same as Fig.~\ref{MSPX_q0_100}, but in a broader range
of $k$.}
\label{MSPX_q0_500}
\end{figure}

Our Fig.~\ref{MSPX_q0_100} is the same as Fig.~\ref{MSPX_qpi_100},
but the splitting is calculated at $\varphi = 0$, instead of
$\pi$; and Fig.~\ref{MSPX_q0_500} is the same as
Fig.~\ref{MSPX_q0_100}, but the data are presented in a broader
range of $k$. The dependences qualitatively resemble those
observed at $\varphi = \pi$, but quantitatively they are, of
course, different.

The power-law index $\gamma$ in the algebraic dependences $\Delta
p \propto \mu^\gamma$ depends on the value of $\varphi$ at which
$\Delta p$ is measured; at $\varphi=0$ one has $\gamma = 1$,
whereas at $\varphi=\pi$ one has $\gamma = 1/2$, see
Ref.~\cite{MG97PhyD}. In our treatment, these values follow directly
from analytical representations of the MA-integrals.

\begin{figure}[th!]
\centering
\includegraphics[width=0.5\textwidth]{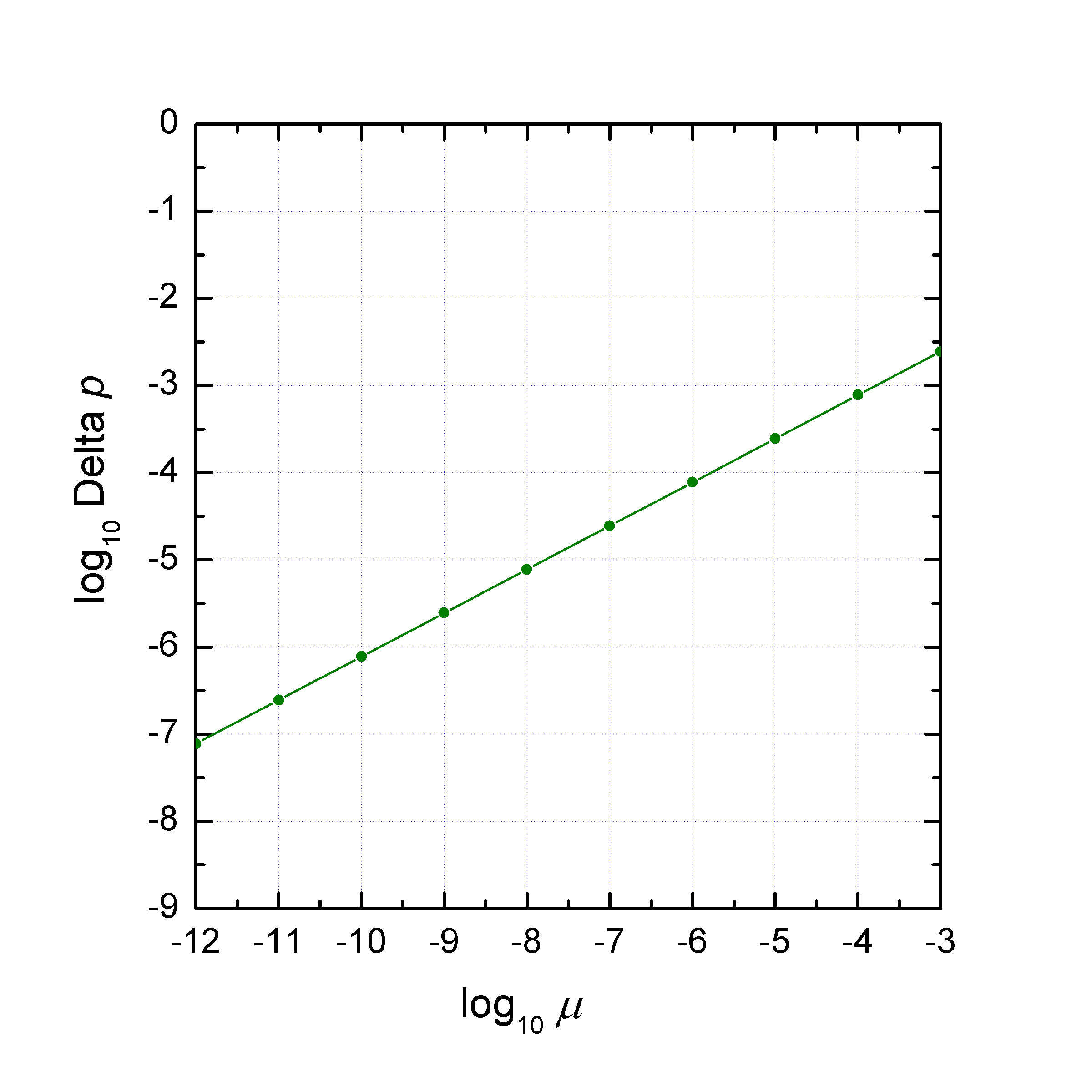} \\
\includegraphics[width=0.5\textwidth]{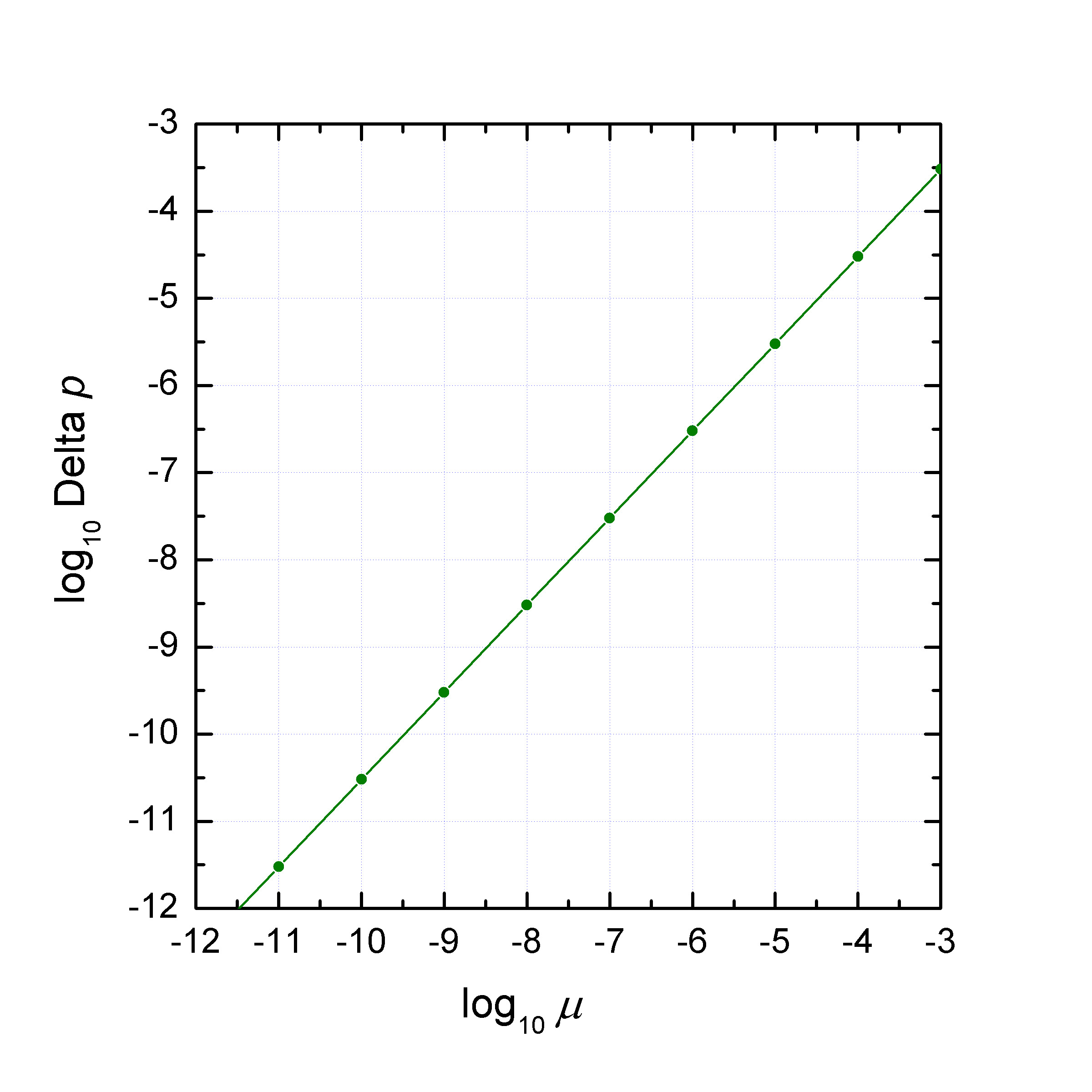} \\
\caption{The separatrix splitting amplitude in
model~(\ref{ex-split})--(\ref{ex-split_2}), as a function of
$\mu$; $\lambda = 100$, $k=\lambda/2 = 50$. Upper panel: the
splitting is calculated at $\varphi = \pi$. Lower panel: the
splitting is calculated at $\varphi = 0$.}
\label{MSPX_mu}
\end{figure}

In Fig.~\ref{MSPX_mu}, the separatrix splitting amplitude for
model~(\ref{ex-split})--(\ref{ex-split_2}), as a function of
$\mu$, is presented at $\lambda = 100$, $k=\lambda/2 = 50$, and
$\epsilon = 10^{-4}$. In the upper panel of Fig.~\ref{MSPX_mu},
the splitting is calculated at $\varphi = \pi$, therefore $\gamma
= 1/2$; this graph is an analytical analogue of figure~4 in
Ref.~\cite{MG97PhyD} and figure~4.7a in Ref.~\cite{Morbi02}; the
visual accordance is excellent. In the lower panel of
Fig.~\ref{MSPX_mu}, the splitting is calculated at $\varphi = 0$,
and therefore $\gamma = 1$.

When constructing theoretical diagrams in this article, we often
used large values of the adiabaticity parameter $\lambda$. However
note that the presented method does not rely on the $\lambda \gg
1$ approximation; mostly condition $\lambda \gtrsim 1$ is implied.
Indeed, if $\lambda \lesssim 1$, then the interacting resonances
strongly overlap, and regular islands, corresponding to secondary
resonances, diminish and disappear, thus making the problem of
determining the sizes of the resonance cells ill-defined.

In the presented figures, the curves have been calculated using
explicit, analytically rather simple formulas. Therefore, there is
no need to check any computational efficiency or accuracy here. On
the other hand, the separatrix splitting was measured in
Ref.~\cite{MG97PhyD} using a numerical computational procedure.
Our analytical curves prove the computational accuracy of the
procedure used in Ref.~\cite{MG97PhyD}.

\section{A paradigm for secondary resonances}
\label{sec_pm}

It is plausible to reformulate the paradigmatic
model~(\ref{ex-split})--(\ref{ex-split_2}) for secondary
resonances so as to introduce symmetry between the dynamics
above and below (in the canonical momentum) the guiding resonance
cell. Indeed, the terms $\propto$~$\cos((k+1)\varphi - t)$ and
$\propto$~$\cos((k-1)\varphi - t)$ in Eq.~(\ref{ex-split_2})
correspond to two island chains situated both above the main
resonance cell: the frequencies $\mathrm{d}\varphi / \mathrm{d}t$
are both positive. Conversely, the island chains will be situated
symmetrically above and below the cell if the signs of $t$ in the
two perturbation terms were different. In this respect, it is
appropriate to take, as a new paradigm, the Hamiltonian with
perturbation terms defined as $\propto$~$\cos(k \varphi - t)$ and
$\propto$~$\cos(k \varphi + t)$.

Therefore, let us consider Hamiltonian (\ref{ex-split}) with
the same ${\cal H}_{\rm res}$ (given by Eq.~(\ref{ex-split_1})), but
with the new perturbation

\begin{equation}
{\cal R}(p,\varphi,t)=-\frac{\epsilon\mu}{2} \left( \cos (k
\varphi+t) +\cos (k \varphi-t ) \right) ,
\label{ex-split_3}
\end{equation}

\noindent  as in Hamiltonian~(\ref{h}) with $a = b$. Then, the
splitting sizes at $\varphi = 0$ and  $\varphi = \pi$, according
to Eqs.~(\ref{Delta_p_w}), are given by

\begin{equation}
\Delta p_{\varphi = 0} \approx \frac{\Delta w}{2 \lambda} =
\frac{1}{\lambda}
W^+(\varepsilon,\lambda,\eta) =
\frac{1}{\lambda} W(\varepsilon,\lambda) ,
\end{equation}

\begin{equation}
\Delta p_{\varphi = \pi} \approx \frac{\sqrt{\Delta w}}{\lambda}
= \frac{1}{\lambda}
( 2 W^+(\varepsilon,\lambda,\eta) )^{1/2} =
\frac{1}{\lambda} ( 2 W(\varepsilon,\lambda) )^{1/2} ,
\end{equation}

\noindent where $W^+$ and $W$ are given by Eqs.~(\ref{samWpmk})
and (\ref{W_k}), and $\eta = 0$.

\begin{figure}[th!]
\centering
\includegraphics[width=0.5\textwidth]{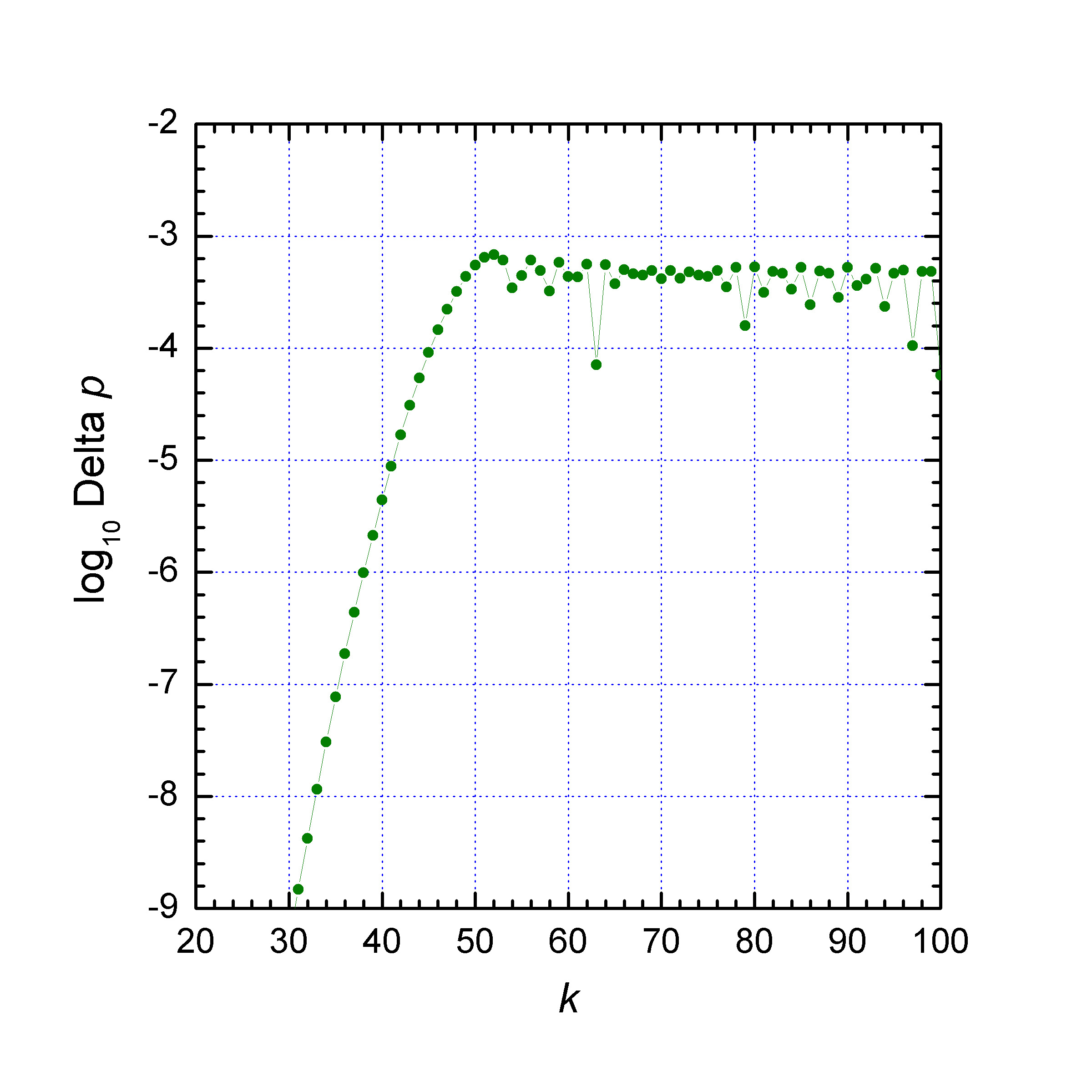} \\
\caption{The separatrix splitting amplitude in
model~(\ref{ex-split_3}) (with $a = b$), as a function of $k$. The
splitting is calculated at $\varphi = \pi$. Note the absence of
any prominent dip similar to that present at $k \sim 70$ in
Fig.~\ref{MSPX_qpi_100}.}
\label{SPX_qpi_100}
\end{figure}

\begin{figure}[th!]
\centering
\includegraphics[width=0.5\textwidth]{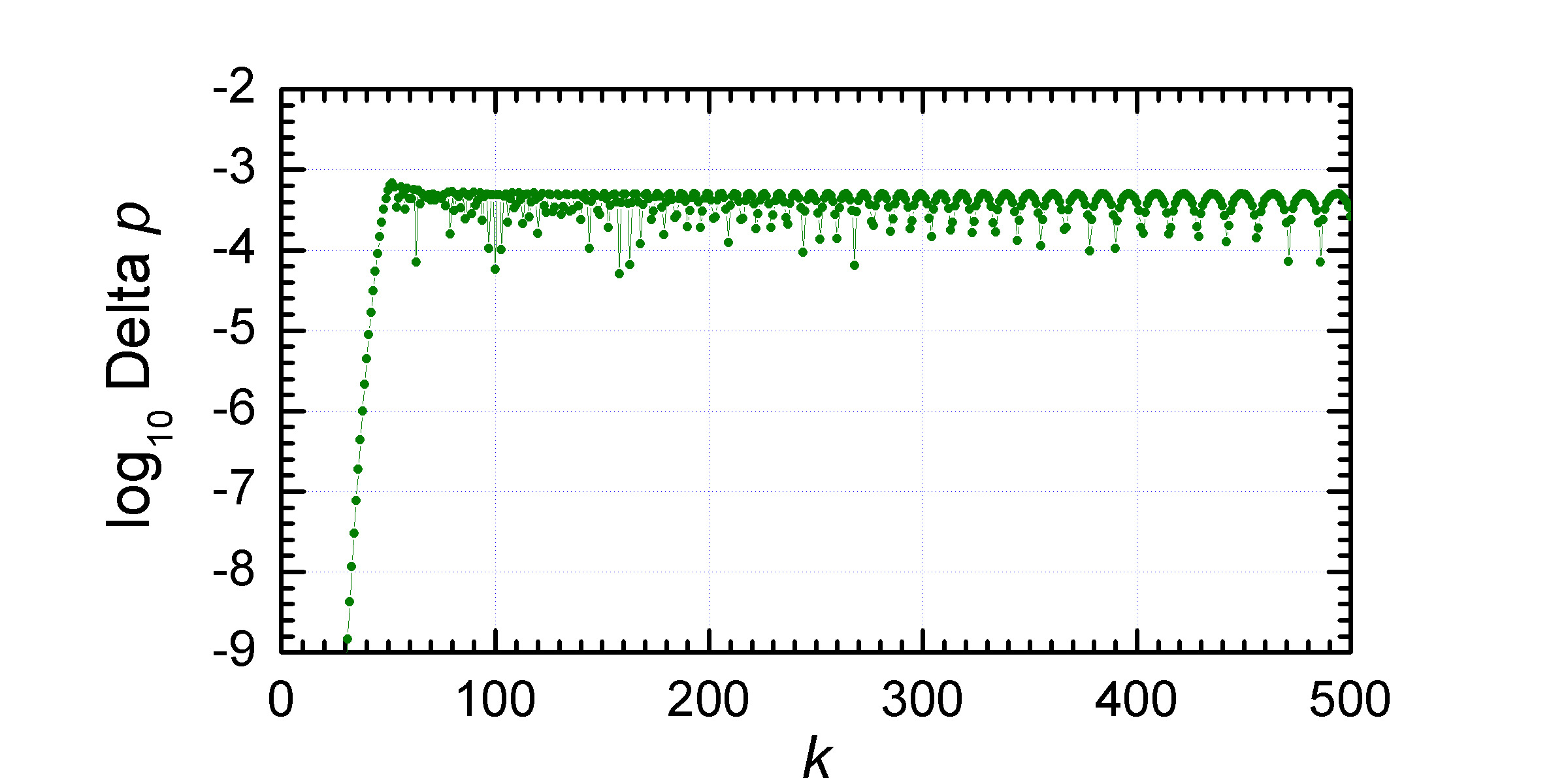} \\
\caption{The same as Fig.~\ref{SPX_qpi_100}, but in a broader range of $k$.}
\label{SPX_qpi_500}
\end{figure}

In the symmetric model~(\ref{ex-split_3}), the separatrix
splitting size is presented, as a function of $k$, in
Fig.~\ref{SPX_qpi_100}. The splitting is calculated at $\varphi =
\pi$. Fig.~\ref{SPX_qpi_500} is the same as
Fig.~\ref{SPX_qpi_100}, but presented in a broader range of $k$.
In Fig.~\ref{SPX_qpi_500}, note the prominent (by an order of
magnitude) quasi-regular ``comb-like'' variations of the
separatrix splitting at large values of $k$ Their nature is the
same as that discussed above in model~(\ref{ex-split_2}). These
variations arise on the attractor of map~(\ref{MA_map}); see
Fig.~\ref{la_100}.

\begin{figure}[th!]
\centering
\includegraphics[width=0.5\textwidth]{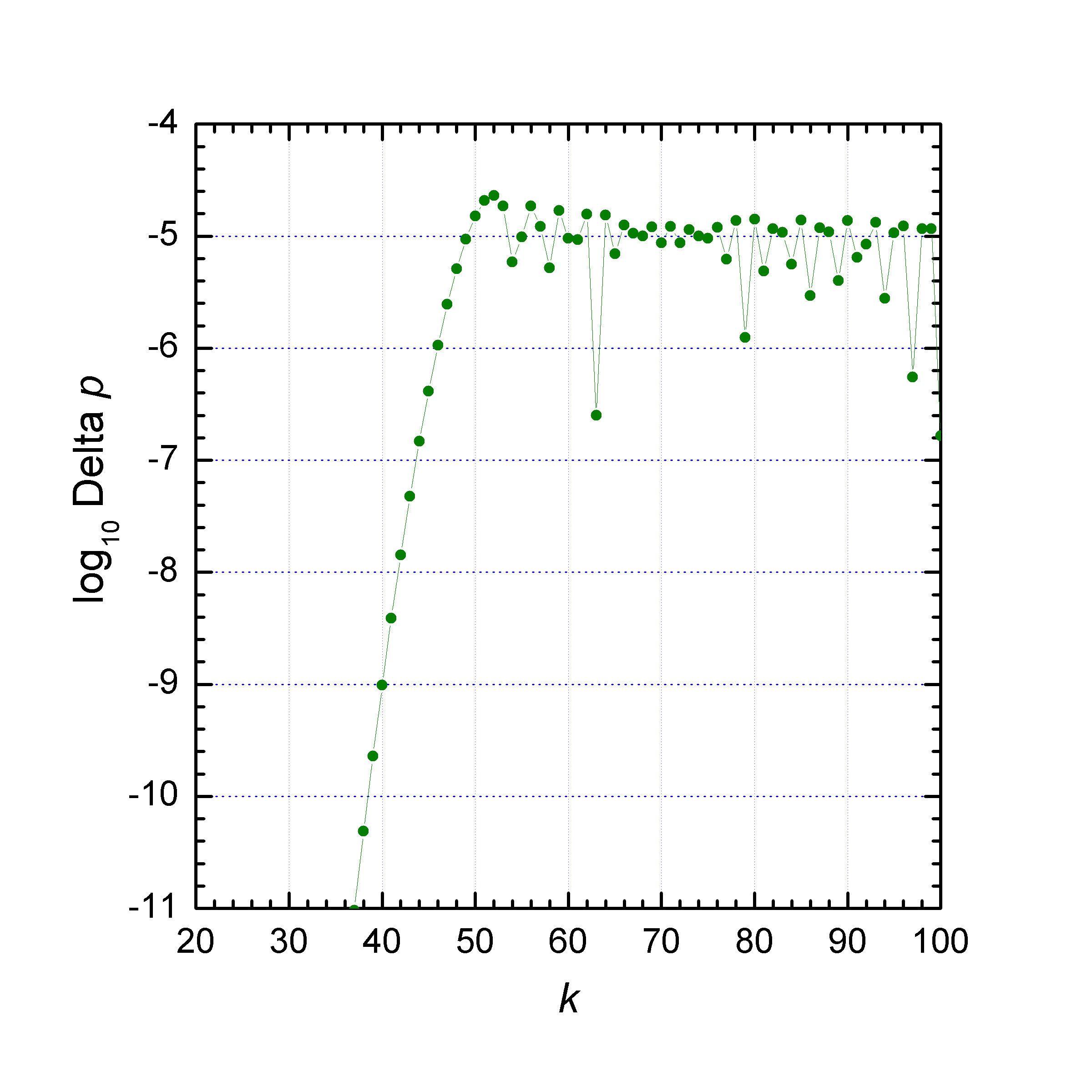} \\
\caption{The same as Fig.~\ref{SPX_qpi_100}, but the splitting is
calculated at $\varphi = 0$.} \label{SPX_q0_100}
\end{figure}

\begin{figure}[th!]
\centering
\includegraphics[width=0.5\textwidth]{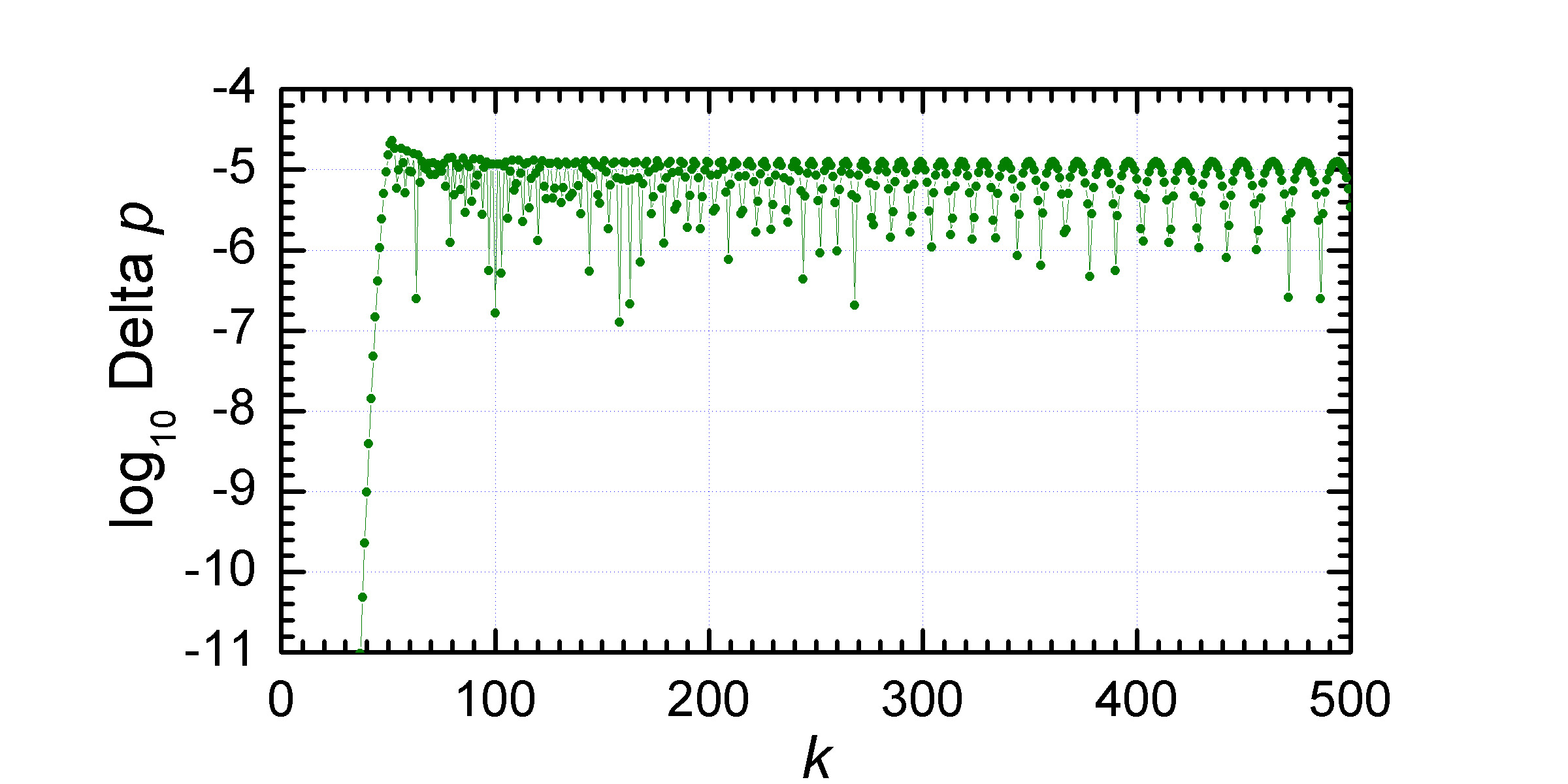} \\
\caption{The same as Fig.~\ref{SPX_q0_100}, but in a broader range
of $k$.}
\label{SPX_q0_500}
\end{figure}

Fig.~\ref{SPX_q0_100} is the same as Fig.~\ref{SPX_qpi_100}, but
the splitting is calculated at $\varphi = 0$, and
Fig.~\ref{SPX_q0_500} is the same as Fig.~\ref{SPX_q0_100}, but
the data are presented in a broader range of $k$.
In the symmetric model~(\ref{ex-split_3}), the qualitative
behaviour of the dependences (the sharp initial rise that changes,
at a critical value of $k$, to a comb-like plateau) is the same as
observed at $\varphi = \pi$, but quantitatively the sizes are
different.

\begin{figure}[th!]
\centering
\includegraphics[width=0.5\textwidth]{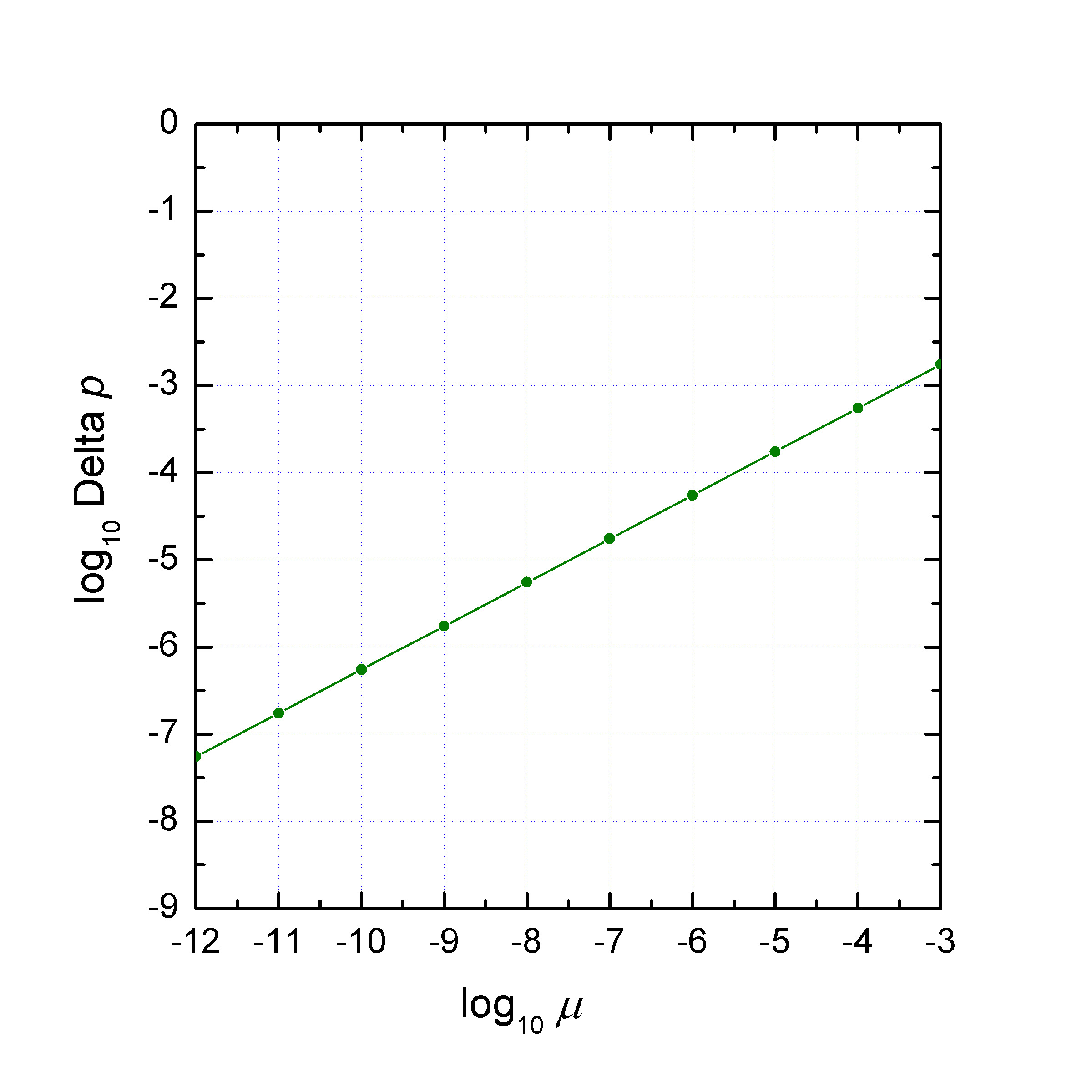} \\
\includegraphics[width=0.5\textwidth]{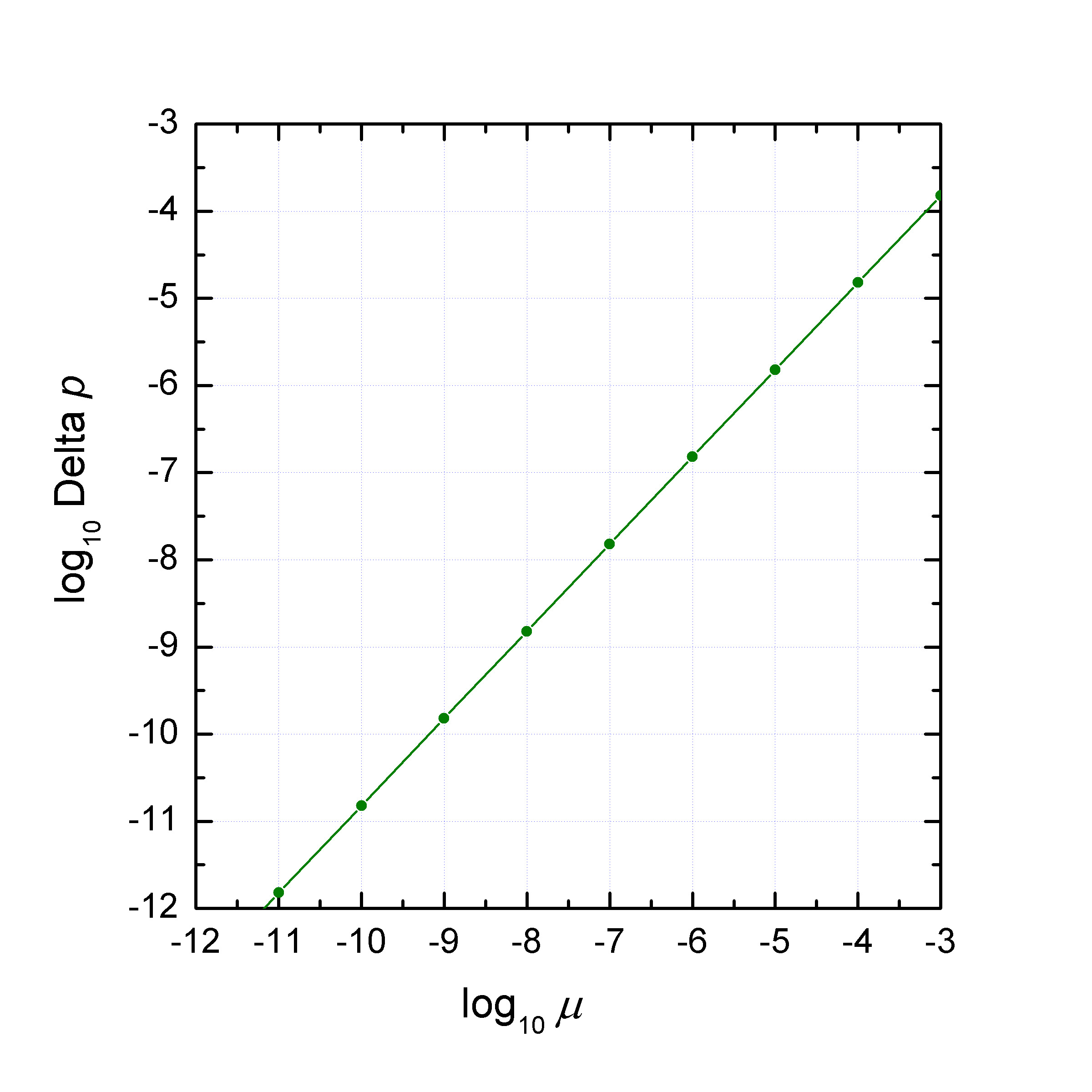} \\
\caption{The separatrix splitting amplitude for the symmetric
perturbation model~(\ref{ex-split_3}), as a function of $\mu$;
$\lambda = 100$, $k=\lambda/2 = 50$. Upper panel: the splitting is
calculated at $\varphi = \pi$. Lower panel: the splitting is
calculated at $\varphi = 0$.} \label{SPX_mu}
\end{figure}

In Fig.~\ref{SPX_mu}, the separatrix splitting size $\Delta p$ in
the symmetric perturbation model~(\ref{ex-split_3}) is presented
as a function of $\mu$; $\lambda = 100$, $k=\lambda/2 = 50$. In
the upper panel, the splitting is calculated at $\varphi = \pi$,
and, in the lower panel, at $\varphi = 0$. The power-law indices
$\gamma$ in the dependences $\Delta p \propto \mu^\gamma$ have
values $\gamma = 1$ at $\varphi=0$ and $\gamma = 1/2$ at
$\varphi=\pi$.

\section{The standard map: resonant normal forms}

A mathematically simple setting of a problem and a simple
formulation of its solution do not often imply that ways of
solving the problem are simple. What is more, arriving at simple
final formulas does not always imply that intermediary analytical
calculations are simple. E.~g., normalizing transformations in
problems on Hamiltonian dynamics often involve complicated
analytical procedures \cite{C79PhR,S08CoPhC}, and an example can
be given, when obtaining some line-sized formulas (for the
coefficients of normal forms) consumed gigabytes of computer
memory \cite{S08CoPhC}.

Here we show how sizes of secondary resonances can be estimated as
a by-product of calculation of MA-integrals, in a rather
straightforward and mathematically simple way. Specifically, we consider
the standard map Hamiltonian~(\ref{h_stm2}) and show how sizes of
secondary resonances in the map's phase space can be estimated in
this new way. The method, as introduced here, is a general one: it
can be straightforwardly used whenever the MA-integrals can be
calculated analytically. Its application to the standard map is
given here as an example, because we can compare the newly derived
normal forms with those obtained already by traditional
normalization.

Before considering this problem, let us recall some necessary data
on the separatrix splitting in the standard map model. The
standard map

\vspace{-6mm}

\begin{eqnarray}
y_{i+1} &=& y_i + K \sin x_i \ \ \ (\mbox{mod } 2 \pi), \nonumber \\
x_{i+1} &=& x_i + y_{i+1} \ \ \ (\mbox{mod } 2 \pi)
\label{stm2} ,
\end{eqnarray}

\noindent describes the motion in an infinite multiplet of
equally-sized equally-spaced resonances, as it is clear from its
Hamiltonian \cite{C79PhR}:

\vspace{-3mm}

\begin{equation}
H = \frac{y^2}{2} + \frac{K}{(2 \pi)^2} \sum_{m=-\infty}^{+\infty}
\cos(x - m t) .
\label{h_stm2}
\end{equation}

\noindent The variables $x_i$, $y_i$ of map~(\ref{stm2})
correspond to the variables $x(t_i)$, $y(t_i)$ of the continuous
system~(\ref{h_stm2}) taken at time moduli $2 \pi$ (see, e.~g.,
Ref.~\cite{C79PhR}).

In this model, Lazutkin obtained \cite{L05JMS} an asymptotic (at
$K \ll 1$) formula for the separatrix splitting angle $\alpha$. At
the first intersection of the separatrices with the line $x =
\pi$, one has

\begin{equation}
\alpha = \frac{\pi}{h^2} \exp \left( -\frac{\pi^2}{h} \right)
\sum_{m=0}^\infty c_m h^{2 m} , \label{alpha}
\end{equation}

\noindent where

\begin{equation}
h = \ln \left(1 + \frac{K}{2} + \left(K + \frac{K^2}{4}
\right)^{1/2} \right) , \label{hsxst}
\end{equation}

\noindent and the first three coefficients $c_m$ are given by the
formulas

\begin{equation}
c_0 = f_0 , \quad c_1 = f_1 - \frac{c_0}{4} , \quad c_2 = f_2 -
\frac{c_1}{4} - \frac{25 c_0}{72} , \label{csxst}
\end{equation}

\noindent where

\vspace{-5mm}

\begin{equation}
f_0 = 1118.82 \dots , \quad f_1 = 18.5989 \dots , \quad f_2 =
-2.17205 \dots , \label{fsxst}
\end{equation}

\noindent see Refs.\cite{L05JMS,G99CMP}.

Using formula~(\ref{alpha}), a correction factor $R_\mathrm{st}$
can be calculated, by which the MA-integrals for the standard map
Hamiltonian should be multiplied:

\begin{equation}
R_\mathrm{st} \approx \frac{1}{16 \pi^3}(c_0 + c_1 h^2 + c_2 h^4)
. \label{Rsxst}
\end{equation}

\noindent Therefore, at $K \ll 1$, one has $h \approx K^{1/2}$,
and the factor $R_\mathrm{st}$ is straightforwardly expressed
\cite{VC98JETP,V99JETP,S20ASSL} through Lazutkin's splitting
constant $f_0 = 1118.82 \dots$, namely, $R_\mathrm{st} = f_0 /(16
\pi^3) \approx 2.2552$.

Next, recall some necessary data on the properties of the
near-separatrix motion, as described by the separatrix map theory.

In the case of weak perturbations ($\varepsilon \ll 1$), the
dynamical behaviour described by the separatrix map~(\ref{sm}) is
symmetric with respect to the sign of the relative energy
$w$.\cite{S98PS} Let $w$ be positive. Let $w^{(k)}$ be the
location of the center of resonance of any integer order $k \geq
1$, and $\Delta w^{(k)}$ be the half-width of secondary resonance
of order $k$. It is known \cite{LL92,S98PS} that, at $\lambda \gg
1$,

\begin{equation}
w^{(k)} = 32 \exp({- 2 \pi k / \lambda}) ,
\label{eq_wm}
\end{equation}

\noindent and $\Delta w^{(k)}$ can be derived by means of
linearization in $w$ of map~(\ref{sm}) near the corresponding
fixed point: considering the Hamiltonian of the approximating
standard map, one finds

\begin{equation}
\Delta w^{(k)} (\varepsilon, \lambda) = 2 \left( {w^{(k)} W
(\varepsilon, \lambda)/ \lambda} \right)^{1/2} ,
\label{Delta_wk}
\end{equation}

\noindent where $W (\varepsilon, \lambda)$ is given by
Eq.~(\ref{W1}); see Ref.~\cite{S98PS}.

With Eqs.~(\ref{Delta_p_w}) and (\ref{Delta_wk}), the resonant
harmonic's size in the canonical momentum $p$, at any $k$th order
and $\lambda \gtrsim 1$, can be found as

\begin{equation}
\Delta p \approx \frac{16 \sqrt{\pi \lambda} }{k} \exp \left( -
\frac{\pi k}{\lambda} - \frac{\pi \lambda}{4} \right) .
\label{dp_method1}
\end{equation}

In what follows, the procedure for calculation of sizes $\Delta p$
based on Eqs.~(\ref{Delta_wk}) and (\ref{dp_method1}) is called
the MA-based method~1. We shall see that this method has a rather
limited field of application.

Note that, within this method, a more precise estimate of the
secondary resonance location $w^{(k)}$ can be obtained by solving
the equation

\begin{equation}
\lambda u {\mathbf K}(u) - \pi k = 0 ,
\label{eq_location}
\end{equation}

\noindent with respect to $u$ numerically, and then evaluating

\begin{equation}
w^{(k)} = \frac{2}{u^2} - 2 .
\end{equation}

\noindent Eq.~(\ref{eq_location}) is derivable from the refined
expression~(\ref{dtau}) for the increment of the phase $\tau$ in
Eq.~(\ref{sm}) at positive values of $w$ (i.~e., in the case of
the resonance phase rotations). However, for our purposes, the
precision provided by formula~(\ref{eq_wm}) is sufficient, and we
use it within method~1 in what follows.

Let us temporarily set the stochasticity parameter $K$ in the
standard map~(\ref{stm2}) and Hamiltonian~(\ref{h_stm2}) equal to
its critical value,\footnote{On the critical $K$ value, see,
e.~g., Ref.~\cite{LL92}.} $K=K_\mathrm{G}=0.971636...$. Then, the
corresponding critical value of the adiabaticity parameter is
$\lambda = 2 \pi/\sqrt{K_\mathrm{G}} = 6.374235...$.

First, we consider secondary resonance with $k=3$. For the
resonance location in the canonical momentum $y$ one has
$y_\mathrm{res} = 1/3$; therefore, the resonance half-width

\begin{equation}
\Delta y = y_\mathrm{res} \frac{\Delta w^{(k)}}{2} = \frac{\Delta w^{(k)}}{6} ,
\end{equation}

\noindent where $k=3$. For the resonance half-width in $w$, one
obtains $\Delta w^{(3)} \approx 0.219$ and $\Delta y \approx
0.036$.

Following Ref.~\cite{C79PhR}, let us define the normalized
stochasticity parameter $\varkappa = K/(4 \pi^2)$ and
$\varkappa_\mathrm{G} = K_\mathrm{G} / (4 \pi^2)$. Using
eq.~(5.21) from Ref.~\cite{C79PhR}, one obtains

\begin{equation}
\Delta y =|S_3|^{1/2} \varkappa_\mathrm{G}^{3/2} ,
\end{equation}

\noindent where

\begin{equation}
S_3 = \sum_{i=-\infty}^{i=+\infty} \sum_{j=-\infty}^{j=+\infty}
\left[ \left(i - \frac{1}{3}\right) \cdot \left(j -
\frac{1}{3}\right) \cdot \left(i + j - \frac{2}{3}\right)^2
\right]^{-1} .
\end{equation}

\noindent Setting $\pm 400$ for the summation limits, instead of
$\pm \infty$, one obtains $S_3 = -86.52...$. Setting the limits equal
to $\pm 300$, one obtains practically the same $S_3$ value
($-86.50...$), which, therefore, we adopt as the true one. Note
that in Ref.~\cite{C79PhR}, a slightly different value (namely,
$-86.4$) was obtained, also by direct numerical summation. We set
$S_3 = -86.5$ (instead of $-86.4$), and obtain $\Delta y =
0.0359$. This is evidently consistent with our MA-based estimate
($\Delta y \approx 0.036$).

\begin{figure}[th!]
\centering
\includegraphics[width=0.5\textwidth]{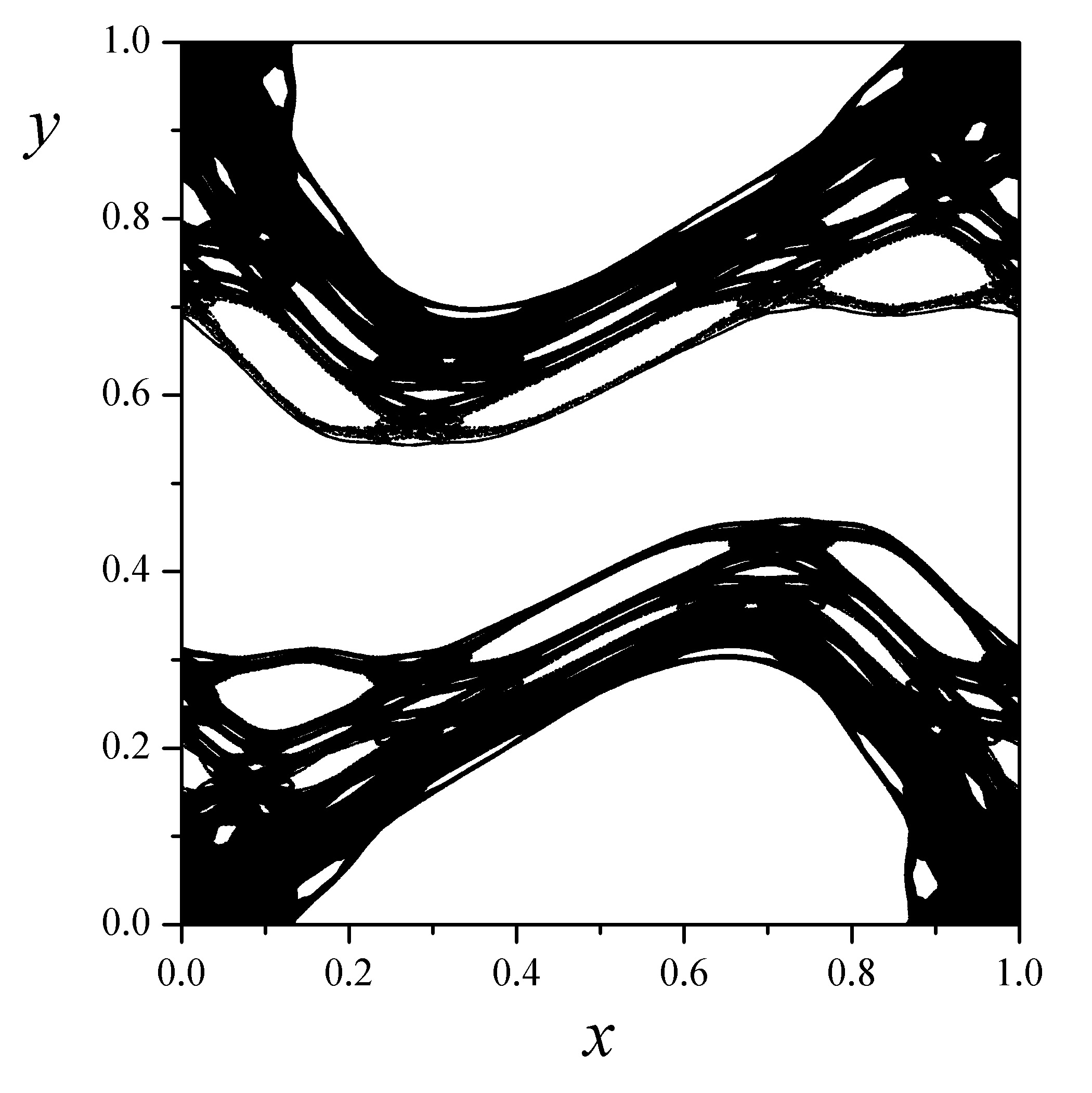} \\
\caption{The phase portrait of the standard map~(\ref{stm2}) at
$K=K_\mathrm{G}=0.971636...$. Solely the chaotic domain around the
integer resonance is shown. Note the 3/1 resonance (the chain of
three islands) at the layer's border, at $y \sim 0.3$ and $\sim
0.7$.} \label{p0p971636}
\end{figure}

On the other hand, from the phase portrait in figure~1 in
Ref.~\cite{Meiss08PJP} (see also our Fig.~\ref{p0p971636}), one
finds the half-width of resonance with $k=3$ to be $\sim 0.03$. A
reasonable accordance with our analytical estimates is clearly
present.

Let us show that the given procedure works out even in the
ultimate\footnote{That with minimum possible $k$.} case $k=2$. For
the resonance location one has $y_\mathrm{res} = 1/2$, and the
resonance half-width

\begin{equation}
\Delta y = y_\mathrm{res} \frac{\Delta w^{(k)}}{2} = \frac{\Delta w^{(k)}}{4} ,
\end{equation}

\noindent where $k=2$. Hence $\Delta w^{(2)} \approx 0.358$, and
$\Delta y \approx 0.089$.

According to equations~(5.22) in Ref.~\cite{C79PhR},

\begin{equation}
\Delta y = \sqrt{ |S_2| } \varkappa = \pi \varkappa ,
\end{equation}

\noindent where $S_2 = -\pi^2$; see equations~(5.15) in
Ref.~\cite{C79PhR}. At $\varkappa = \varkappa_\mathrm{G}$, one has
$\Delta y = 0.0773$. On the other hand, from the standard map
phase portrait, constructed in figure~1 in Ref.~\cite{Meiss08PJP}
for the same $\varkappa$ value, one finds the resonance half-width
at $k=2$ as $\approx 0.06$. We see that, even in the ultimate case
of $k=2$, all our three estimates for the half-width are in
reasonable accordance.

However, the given MA-based method is not applicable at all values
of $K$, but solely at $K \sim 1$. Indeed, for the resonance
location $w^{(k)}$ (with respect to the unperturbed separatrix) to
be less than unity (so that the separatrix map description is
valid), one should have $k \gtrsim \lambda/2$, as follows from
Eq.~(\ref{eq_wm}). On the other hand, normal forms can be
constructed only up to the optimal order $k \approx
\lambda/2$,\cite{MG97PhyD} i.~e., one should have $k \lesssim
\lambda/2$. Therefore, solely at $k \sim \lambda/2$ one expects
the procedure to be valid. If $k=3$, this condition is satisfied
at $\lambda \sim 6$, i.~e., at $K \sim 1$; and, if $k=2$, it is
satisfied at $\lambda \sim 4$, also close to the critical
condition; that is why method~1 produces adequate results in these
two cases, $k=2$ and $k=3$.

Now let us propose and consider another method, also MA-based (we
call it method~2), which turns out to have a much broader field of
application than method~1. It is based on the assumption that any
secondary resonance produces the same splitting of the
separatrices of the guiding resonance as the original perturbing
resonance provides; in this sense, it has the same ``splitting
strength.'' Formally, by the splitting strength of any perturbing
resonance we imply the amount of splitting, caused by this
resonance, of the separatrix of the guiding resonance. The given
assumption is plausible, because the interaction properties of any
resonances should not depend on the choice of the phase space
domain considered.

Then we can derive the sizes of secondary resonances by equating
their splitting strengths to that of the perturbing resonance that
is explicitly present in the original non-normalized Hamiltonian.

Therefore, any $k$th-order secondary resonance (present as two
symmetric perturbing terms in model Hamiltonian~(\ref{h})) is,
first of all in this method, characterized by its splitting
strength $D_k$. By definition, $D_k$ is equal to $W$ at the given
value of $k$. According to Eq.~(\ref{W_k}) at $\lambda \gtrsim 1$,
one has

\begin{equation}
D_k = W \approx \varepsilon \lambda A_{2k}(\lambda) ,
\end{equation}

\noindent where $k = 1, 2, ...$. To estimate $A_i(\lambda)$ (with
$i = 1, 2, \dots$), we use the following asymptotic
representation:

\begin{equation}
A_i(\lambda) \approx \frac{4 \pi}{(i-1)!} (2 \lambda)^{i-1}
\exp \left( - \frac{\pi \lambda}{2} \right) ,
\label{eqC77}
\end{equation}

\noindent valid if $i \ll \lambda$ (see eq.~(A.10) in Ref.~\cite{C79PhR}).

For the standard map Hamiltonian, the adiabaticity parameter
$\lambda=1/ \sqrt{\varkappa}$ (see Eq.~(\ref{h_stm2})), and, at
$k=1$, one has $\varepsilon_1 = 1$. Therefore,

\begin{equation}
D_1 = W_1 \approx \varkappa \lambda A_2(\lambda) \approx 8 \pi
\exp \left( - \frac{\pi}{2 \varkappa^{1/2}} \right) .
\end{equation}

\noindent At $k=2$
one has

\begin{equation}
D_2 = W_2 \approx \varepsilon_2 \frac{16 \pi}{3 \varkappa^2} \exp
\left( - \frac{\pi}{2 \varkappa^{1/2}} \right) ,
\end{equation}

\noindent where $\varepsilon_2$ is the size of the harmonic with
$k=2$; this is the quantity that is sought to be found.

At $k=3$ one has

\begin{equation}
D_3 = W_3 \approx \varepsilon_3 \frac{16 \pi}{15 \varkappa^3} \exp
\left( - \frac{\pi}{2 \varkappa^{1/2}} \right) ,
\end{equation}

\noindent where $\varepsilon_3$ is the sought size of the harmonic
with $k=3$.

In this way one finds $D_k$ at any $k$. Then, solving the
equations

\begin{equation}
D_1 = D_2, \quad D_1 = D_3, \quad ... , \quad D_1 = D_k, \quad ... ,
\end{equation}

\noindent one finds the sizes $\varepsilon_k$ at any $k$.

At $k=2$, the solution of the equation $D_1=D_2$ is

\begin{equation}
\varepsilon_2 = \frac{3}{2} \varkappa^2 .
\end{equation}

\noindent To take into account the specific properties of the
standard map (namely, $\varepsilon$ is not small, as it is equal
to 1, and the number of explicit perturbing resonances is
infinite, as given by Eq.~(\ref{h_stm2})), the separatrix
splitting size should be multiplied by a correction factor, see
Refs.~\cite{C79PhR,S20ASSL}. Therefore, one should multiply
$\varepsilon_2$ by this correction factor, namely,
$R_\mathrm{st}=2.2552...$; see Eq.~(\ref{Rsxst}). One arrives at

\begin{equation}
\varepsilon_2' \approx 3.38 \, \varkappa^2 . \label{eps2prim}
\end{equation}

\noindent At the next order ($k=3$), by solving the equation
$D_1=D_3$ we find

\begin{equation}
\varepsilon_3 = \frac{15}{2}\varkappa^3 .
\end{equation}

\noindent When multiplied by $R_\mathrm{st}=2.2552...$,
$\varepsilon_3$ becomes

\begin{equation}
\varepsilon_3' \approx 16.9 \, \varkappa^3 . \label{eps3prim}
\end{equation}

For the standard map Hamiltonian, the quantity $\varepsilon_k$
(that is the amplitude
of the separatrix of the $k$th-order resonant harmonic) is related
to the harmonic's size $\Delta y_k$, measured in the canonical
momentum $y$, by the formula

\begin{equation}
\Delta y_k = 2 \sqrt{\varepsilon_k} .
\label{eq_dy_dw}
\end{equation}

\noindent This relation follows from the separatrix equation for
the $k$th-order harmonic. Indeed, generalizing
equations~(5.16)--(5.22) in Ref.~\cite{C79PhR}, one can write down
the $k$th-order harmonic's Hamiltonian in the form

\begin{equation}
H_k = \frac{y^2}{2} - {\cal F}_k \cos ( k \varphi - t ) ;
\label{H_k}
\end{equation}

\noindent therefore, the separatrix is given by

\begin{equation}
y_k = \frac{1}{k} \pm 2 \sqrt{{\cal F}_k} \cos \left( \frac{k}{2}
\varphi - \frac{t}{2} \right) .
\label{p_k}
\end{equation}

\noindent Particular versions of these formulas (at $k=2$ and
3) are presented as equations~(5.16)--(5.22) in Ref.~\cite{C79PhR}.

Using general formula~(\ref{eq_dy_dw}), from our
Eqs.~(\ref{eps2prim}) and (\ref{eps3prim}) we find $\Delta y_2
\approx 2 \sqrt{\varepsilon_2'} \approx 3.68 \varkappa$ at $k=2$,
and $\Delta y_3 \approx 2 \sqrt{\varepsilon_3'} \approx 8.23
\varkappa^{3/2}$ at $k=3$. These relations can be compared to
those derived in Ref.~\cite{C79PhR} by a direct normalization of
the Hamiltonian (see equation~(5.22) in Ref.~\cite{C79PhR}):
$\Delta y_2 = \pi \varkappa$ at $k=2$, and $\Delta y_3 \approx
9.30 \varkappa^{3/2}$ at $k=3$. Obviously, our method~2 and the
direct normalization provide identical results for the scalings in
$\varkappa$ and rather similar values for the coefficients of the
scalings.

Note that any further convincing comparison in analytical accuracy
is not needed here, because the obtained analytical forms are all
the same as obtained by the traditional normalization; solely the
numerical coefficients can be somewhat different,
because the nature of approximation is different.

Concerning the calculational efficiency and computational
accuracy, the proposed method does not involve any programmable
computations (broadly used in the traditional normalization): the
sizes of resonances are obtained directly, using formulas, which
are explicit and rather simple.

\section{Discussion}

The results obtained above by the MA-based method, as applied to
the standard map, are all compiled in Table~1, along with the
results obtained in Ref.~\cite{C79PhR} within the same problem, by
the traditional direct normalization. Within the Table, the
results can be straightforwardly cross-compared. Where accessible,
approximate estimates based on direct inspections of phase
portraits of the standard map are also given.

The columns of this Table are designated as follows: $k$ is the
order of the harmonic; $\Delta y^\mathrm{2nd}$ is the MA-based
(obtained by our method~2, as defined above) size $\Delta y$;
$\Delta y^\mathrm{Ch}$ is $\Delta y$ from Ref.~\cite{C79PhR};
$\Delta y^\mathrm{1st}_{K=K_\mathrm{G}}$ is the MA-based (obtained
by our method~1) $\Delta y$ at $K=K_\mathrm{G}$; $\Delta
y^\mathrm{2nd}_{K=K_\mathrm{G}}$ is $\Delta y^\mathrm{2nd}$
calculated at $K=K_\mathrm{G}$; $\Delta
y^\mathrm{Ch}_{K=K_\mathrm{G}}$ is $\Delta y$ at $K=K_\mathrm{G}$,
as obtained in Ref.~\cite{C79PhR} by direct normalization; $\Delta
y^\mathrm{graph}_{K=K_\mathrm{G}}$ is the estimate based on
figure~1 in Ref.~\cite{Meiss08PJP}.

\begin{table}[h]
\textbf{Table 1. Sizes of secondary resonances of the standard map} \\
\medskip
\begin{center}
\begin{tabular}{lllllll}
\hline \hline \\
$k$ & $\Delta y^\mathrm{2nd}$ & $\Delta y^\mathrm{Ch}$ &
$\Delta y^\mathrm{1st}_{K=K_\mathrm{G}}$ &
$\Delta y^\mathrm{2nd}_{K=K_\mathrm{G}}$ &
$\Delta y^\mathrm{Ch}_{K=K_\mathrm{G}}$ &
$\Delta y^\mathrm{graph}_{K=K_\mathrm{G}}$ \\
\hline \\
2 & $3.68 \varkappa$        & $\pi \varkappa$        & 0.089  & 0.091 & 0.077 & $\sim$0.06 \\
3 & $8.23 \varkappa^{3/2}$  & $9.30 \varkappa^{3/2}$ & 0.036  & 0.032 & 0.036 & $\sim$0.03 \\
4 & $26.65 \varkappa^2$     & ---                & 0.0167 & 0.0161 & ---   & $\sim$0.01 \\
5 & $113.1 \varkappa^{5/2}$ & ---                & 0.0102 & 0.0107 & ---   & ---        \\
\hline \hline
\end{tabular}
\end{center}
\end{table}

The data in Table~1 are given up to order $k=5$. Note that
$\Delta y^\mathrm{2nd}$ and $\Delta
y^\mathrm{1st}_{K=K_\mathrm{G}}$ can be easily calculated up to
any order $k$.

Upon inspecting Table~1, we see that the data in the 2nd and 3rd
columns ($\Delta y^\mathrm{2nd}$ and $\Delta y^\mathrm{Ch}$) are
in good accordance. Our method~2 (giving $\Delta y^\mathrm{2nd}$)
and the direct normalization (giving $\Delta y^\mathrm{Ch}$)
provide identical results for the scalings in $\varkappa$ and
similar results for the coefficients of the scalings; however,
using the direct normalization, it is rather difficult to
analytically obtain $\Delta y^\mathrm{Ch}$ even in the lowest
orders of normalization, and the difficulties sharply rise with
order $k$; see the procedure in Ref.~\cite{C79PhR}. In contrast,
our method~2 allows one to easily find sizes of secondary
resonances up to any order $k$.

The sizes of resonances at $K=K_\mathrm{G}$ (given in the last
four columns of the Table), are all evidently consistent with each
other, including the estimates obtained by our method~1.

Note that the graphical (very approximate) estimates from the
phase portraits are provided here solely for auxiliary comparison;
they are not exact enough, but they at least show that these
graphical data agree with our analytical estimates.

Finally, in view of the obtained results, what can be said on the
dynamical essence of the MG-prescription? One may hypothesize that
the MG-prescription $k \sim \lambda / 2$ reflects some radical
change in the behaviour of the secondary resonances at this value
of $k$. Perhaps, they start to merge with the main chaotic layer?
However, this is not so; here we show that the merging occurs at
much greater values of $k$, namely, at $k \sim \lambda^2 / 4$.

An equivalent form of Eqs.~(\ref{sm}), used, e.~g., in in
Refs.~\cite{CS84PhyD,S20ASSL}, is given by

\vspace{-3mm}

\begin{eqnarray}
     y_{i+1} &=& y_i + \sin x_i, \nonumber \\
     x_{i+1} &=& x_i - \lambda \ln \vert y_{i+1} \vert + c
                   \ \ \ (\mbox{mod } 2 \pi),
\label{sm1}
\end{eqnarray}

\noindent where $y = w / W$, $x = \tau + \pi$; and the parameter

\vspace{-3mm}

\begin{equation}
c = \lambda \ln \frac{32}{\vert W \vert}. \label{c}
\end{equation}

\noindent Let $w_\mathrm{bord}$ be the location of the border of
the main chaotic layer, as described by map~(\ref{sm}). It is
known that, if $\lambda \gtrsim 1$, then $w_\mathrm{bord} \approx
\lambda W$ \cite{CS84PhyD,C91CSF}, i.~e., for the layer described
by map~(\ref{sm1}), $y_\mathrm{bord} \approx \lambda $.

Setting the increment in $x$ to be equal to $2 \pi k$ at the
layer's border, one derives, from the second equation of
Eqs.~(\ref{sm1}), the following relation for the order
$k_\mathrm{marg}$ of the marginal (that merging with the main
chaotic layer) integer resonance:

\begin{equation}
\lambda \ln \frac{32}{\lambda W } = 2\pi k_\mathrm{marg} .
\end{equation}

\noindent Eq.~(\ref{W1}) at $\lambda \gtrsim 1$ gives

\begin{equation}
W \approx 8 \pi \varepsilon \lambda^2 \exp \left( -\frac{\pi \lambda}{2} \right) ,
\end{equation}

\noindent and one arrives at the relation

\begin{equation}
\lambda \ln \frac{4}{\pi \varepsilon \lambda^3 } + \frac{\pi
\lambda^2}{2} = 2\pi k_\mathrm{marg} .
\end{equation}

\noindent Asymptotically at $\lambda \gg 1$, it reduces to

\begin{equation}
k_\mathrm{marg} \sim \frac{\lambda^2}{4} .
\end{equation}

\noindent As defined above, the MG-prescription for the optimal
normal form is $k \approx \lambda / 2$. Therefore, the secondary
resonances start to overlap at much higher values of $k$ than that
given by the MG-prescription.

\section{Conclusions}

In this article, we have explored how calculations of the
MA-integrals may provide ways to obtain resonant normal forms for
a given Hamiltonian system. The study has been accomplished in the
framework of the perturbed first fundamental (pendulum) model of
nonlinear resonance.

Within the standard map model, as a paradigmatic example, we have
shown how the newly developed MA-based procedure allows one to
estimate the sizes of secondary resonances of any order (up to the
order of the optimal normal form), without relying on the
cumbersome traditional normalization procedure.

The data on sizes of secondary resonances, obtained by our
method~2, are all consistent with the data obtained by direct
normalization in Ref.~\cite{C79PhR}. Our method~2 and the direct
normalization provide identical results for the scalings in the
stochasticity parameter $\varkappa$ and similar results for the
coefficients of the relations.

However, using direct normalization, it is analytically difficult
to perform even lowest orders of normalization, and the
difficulties sharply rise with order $k$; see the procedure in
Ref.~\cite{C79PhR}. In contrast, our method~2 allows one to easily
estimate sizes of secondary resonances up to any order $k$.

\bigskip


The author is grateful to the referees for useful remarks and
comments.

\bigskip

The author declares no conflict of interest.

\newpage

\end{document}